# A Hierarchical Method for Environmental Sounds Recognition in Complex Acoustic Scenes

Zhouwen Li1, Haijun Lin1*, Bowen Lv1, Yupeng Chen1
1 Hunan Normal University School of Engineering and Design Changsha, China

**Abstract:** Environmental noise identification is essential for noise control, yet conventional single-stage classifiers suffer from class confusion under imbalanced multiclass conditions. This paper proposes a lightweight hierarchical model integrating a Frequency-Time Adaptive Network (FTANet) with a two-stage classification mechanism. FTANet employs parallel asymmetric one-dimensional convolutions to capture temporal dynamics and frequency-band distributions, while adaptive fusion attention (AFA) enhances discriminative acoustic features. In the first stage, FTANet classifies sounds into five coarse-grained categories covering 25 fine-grained classes. In the second stage, samples are routed to the corresponding category-specific classifiers for fine-grained recognition. On a self-constructed dataset, FTANet achieves 95.45% coarse-grained accuracy with only 0.2004M parameters. On UrbanSound8K, it achieves 79.24% mean accuracy and 80.37% mean F1-score in ten-fold evaluation, outperforming compared convolutional models. Field tests on the Tianjin Online Acoustic Environment Monitoring Platform yield 87.94% coarse-grained accuracy and 97.21% conditional fine-grained accuracy for natural sounds correctly routed to C2. The complete hierarchical model achieves 87.18% overall accuracy and 87.40% weighted-average F1-score, demonstrating its feasibility for urban environmental noise monitoring.

## 1. Introduction

With the development of urban transportation, industry, and social activities, the urban sounds are becoming increasingly complex. Different sounds overlap in time, space, and frequency, which has higher requirements for urban sound environment monitoring, public noise control, ecological environment assessment, etc. Conventional acoustic monitoring mainly relies on sound pressure level indicators to characterize environmental noise. Carratù et al. [1] developed a sound level meter capable of automatically estimating measurement uncertainty, while Zagubień and Wolniewicz [2] investigated environmental noise measurement under different microphone positioning conditions. Although these studies have contributed to the reliability of acoustic measurements, sound pressure level indicators alone cannot identify the specific sources of environmental noise. Automatic sound-source identification can therefore complement conventional acoustic measurements by providing information for noise-source analysis, targeted monitoring, and urban noise management.

Environmental sound classification (ESC) has become an important field in acoustic research [3]. and has been widely applied in urban acoustic environment monitoring [4-6], public safety early warning [7], smart-city operations [8], ecological and environmental assessment [9], and surveillance systems [10-12]. Milivojčević et al. [13] investigated vehicle fuel-type classification using psychoacoustic features, while Lv et al. [14] developed a sound recognition method based on spectrogram features and a fusion classification model for animal acoustic monitoring. Wang and Gao [15] employed acoustic-pressure information fusion and a one-dimensional convolutional neural network for pipeline leak detection. These studies demonstrate the value of acoustic feature extraction and intelligent classification in measurement applications. However, environmental sounds are typically nonstationary [16] and frequently affected by background noise and overlapping sources, making reliable multiclass recognition challenging.

Many existing ESC methods employ two-dimensional convolutional neural networks (2D CNNs) to extract features from audio spectrograms [17–20]. However, the time and frequency axes of spectrograms represent different acoustic characteristics: the time axis describes the temporal evolution of sound energy, whereas the frequency axis reflects its spectral distribution [21]. Conventional square convolution kernels jointly process these two dimensions without explicitly distinguishing their characteristics, potentially introducing redundant features. Recent studies in *Measurement* have explored more effective acoustic feature modeling. Li et al. [22] employed time-frequency analysis and an attention mechanism for train noise prediction, while Li et al. [23] combined time-frequency convolution and self-attention for nonstationary mechanical sound-source separation. Liu et al. [24] integrated spectral subtraction with a CNN-TCN model for acoustic fault diagnosis under noisy conditions. Although these approaches demonstrate the importance of time-frequency feature extraction, developing compact models that efficiently capture discriminative temporal and spectral information remains an important issue for environmental sound recognition. Moreover, deeper convolutional architectures may increase computational costs and restrict practical deployment [25].

Another challenge arises from the multiclass structure of environmental sounds. Existing ESC systems commonly employ single-stage classifiers that directly distinguish all fine-grained sound categories. When the number of categories increases, acoustic similarity and class imbalance can lead to classification confusion, particularly among underrepresented classes. Although hierarchical classification has previously been investigated for environmental noise recognition [26], the integration of lightweight time-frequency feature modeling with hierarchical multiclass recognition remains insufficiently explored, particularly for practical urban acoustic monitoring applications.

To solve these problems of time-frequency feature modeling and multiclass classification, this paper proposes a lightweight hierarchical environmental sound recognition method with complex acoustic scenes. The main contributions are written as follows:

1) A lightweight Frequency-Time Adaptive Network (FTANet) is designed. The asymmetric one-dimensional convolutions are employed to extract temporal dynamics and frequency-band distribution features separately, and the adaptive fusion attention mechanism is used to recalibrate the features, while reducing the number of model parameters and enhancing the feature expression ability under complex background noise.

2) A hierarchical classification method is constructed. On the basis of the first level classification results, the overall 25-class classification task is decomposed into several smaller within-category classification tasks, thereby alleviating inter-category confusion under imbalanced data conditions.

## 2. Environmental Sound Recognition Method Based on FTANet and Hierarchical Strategy

Environmental sounds are diverse in types , and it is difficult to require the sounds samples. In addition to, the recognition accuracy of the conventional flat and single-stage classifiers is generally low. This paper proposes an environmental sound recognition model based on FTANet and a two-stage hierarchical classification architecture, which significantly improves the recognition accuracy.

### 2.1. Network Architecture

The architecture of this proposed model for complex environmental sound recognition is shown in Fig. 1. The model framework consists primarily of three components: the input preprocessor, the FTANet based feature model, and the two-stage hierarchical classification. After adaptive framing and Mel/FBank feature extraction of sound signals, a three channel time-frequency representation is formed and input into the FTANet backbone network. Through time-frequency feature extraction, adaptive fusion attention, residual learning, and multi-stage feature modeling, a shared acoustic feature representation is obtained. Based on the first-stage classification result, the corresponding second-stage sub-classifier is then selected to perform within-category fine-grained sound recognition.

FTANet is the core network. Its time-frequency convolution module separately extracts dynamic variations along the time dimension and energy distributions along the frequency dimension. The adaptive fusion attention module further performs channel-wise recalibration of the fused features, thereby enhancing effective features associated with the target sound. On this basis, the two-stage hierarchical classification method decomposes the global multiclass recognition task into coarse-grained category discrimination and within-category fine-grained recognition, which reduces the decision space of the classifiers and minimize the feature interferences between different acoustic categories.

To illustrate the characteristic differences of environmental sounds in time and frequency dimensions, six representative environmental sounds are selected from five coarse-grained categories for analysis, including the bird sound and the frog sound from natural sounds, the blower noise from industrial sounds, the broadcast sound from social activity noises, the automobile engine sound from transportation noises, and the explosion sound from construction noises. By comparing their waveforms, short-time energies, time-frequency spectrograms, and variations in spectral features, it is provided a basis for the architectural design of the Frequency-Time Adaptive Network.

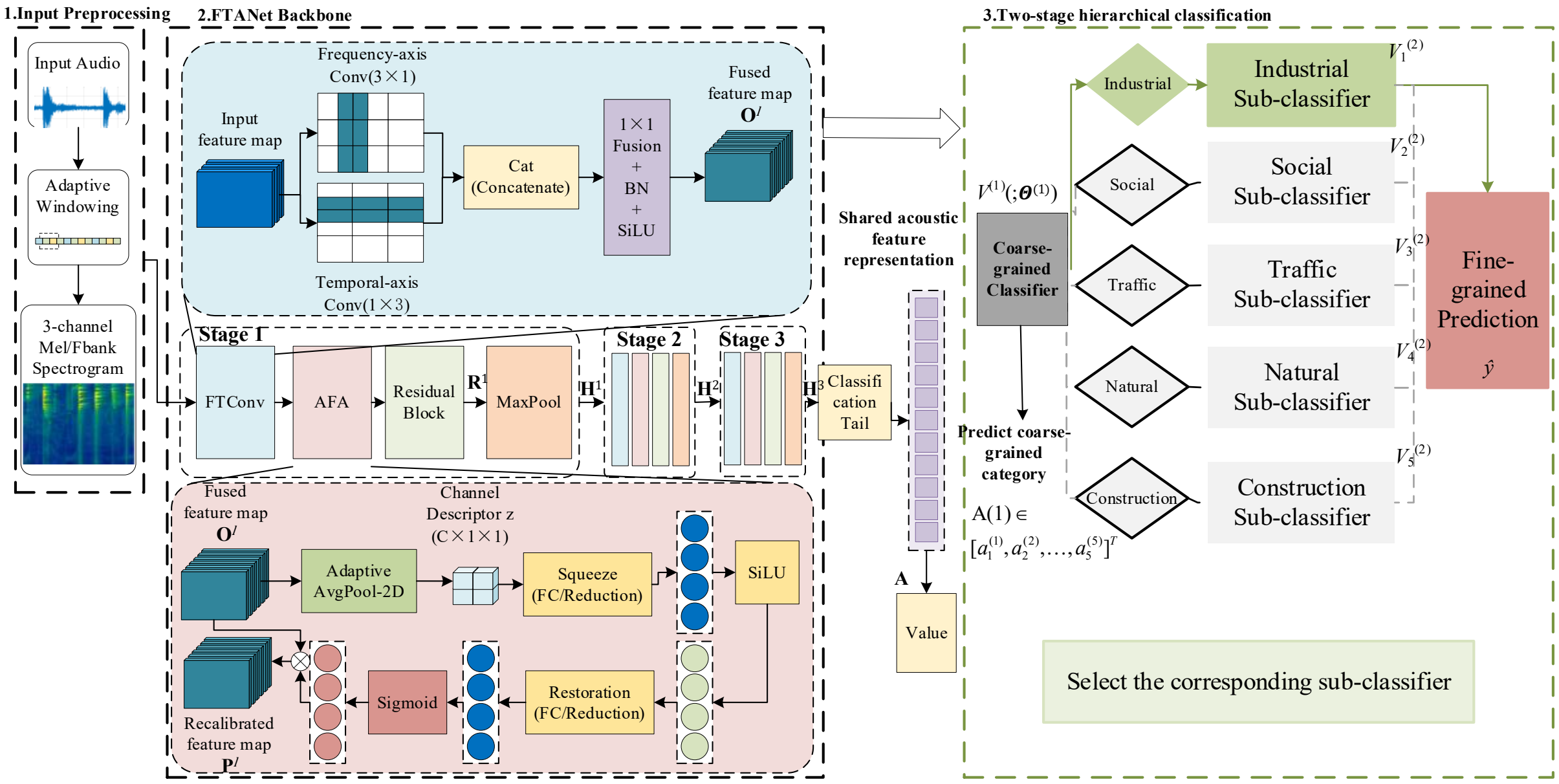

**Fig. 1**. Architecture of this proposed method for feature-modeling and hierarchical classification

## 2.2. Frequency-Time Adaptive Network (FTANet)

Environmental sounds have different time-frequency characteristics.The time-domain and frequency-domain acoustic characteristics of bird calls, vechicle engine noise, blasting noise, blower noises, broadcast sounds, and frog calls are shown in Fig.2-3, respectively. As shown in Fig. 2, The "blasting noise" is manifested as a typical and localized discontinuous pulse peak in the short-term energy curve, with its short-term energy rapidly increasing and decreasing within a local time range, but the blower noise presents a relatively stable energy envelope in short-term energ

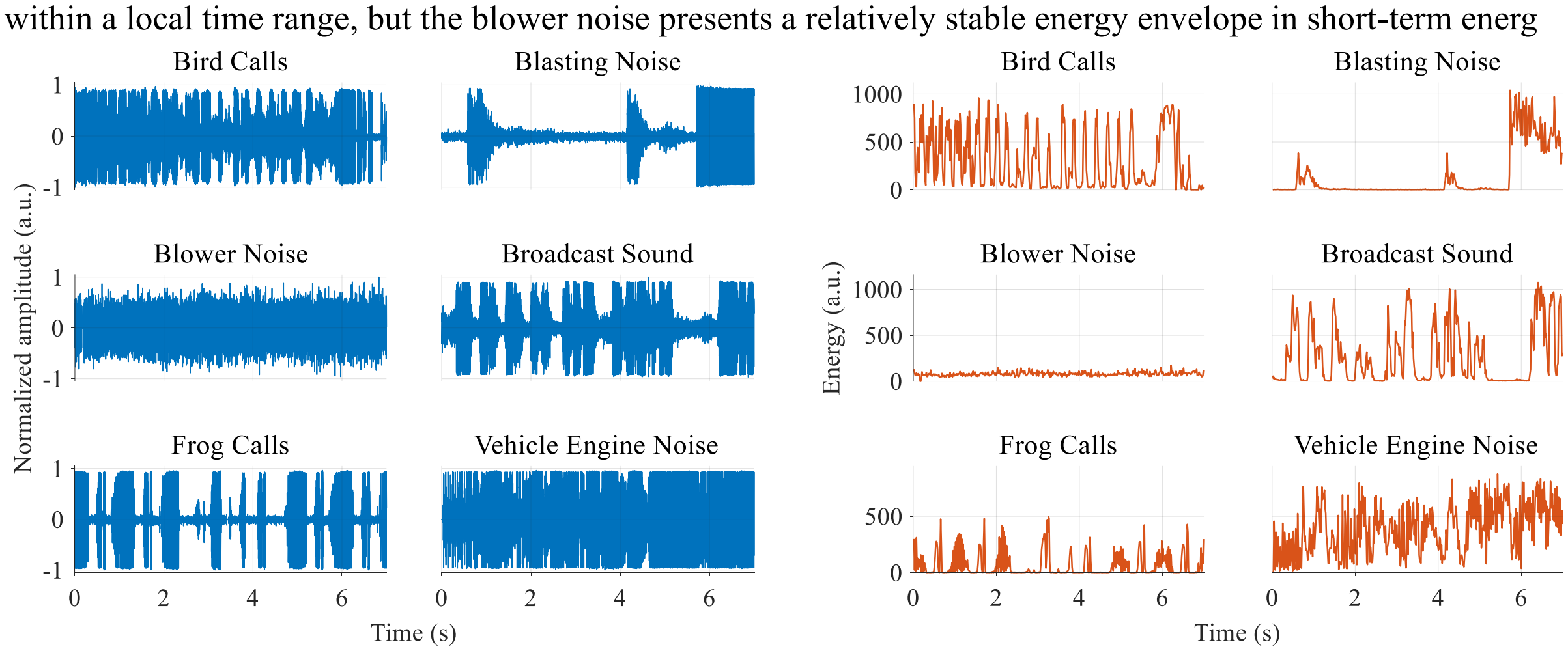

(a) Time-domain plots of the original waveforms (b) Short-time energy curves

**Fig. 2**. Time-domain analysis of the six representative sounds

The differences in frequency dimension between different sounds are shown in Fig.3. The time-frequency spectra of "bird calls" and "frog calls" both show obvious transverse narrowband harmonic spectral lines, with their spectral centroids undergoing high-frequency jumps over time, while the spectral roll off points of " vechicle engine noise" and "blower noise" remain in the low-frequency range for a long time, exhibiting wideband steady-state focusing.

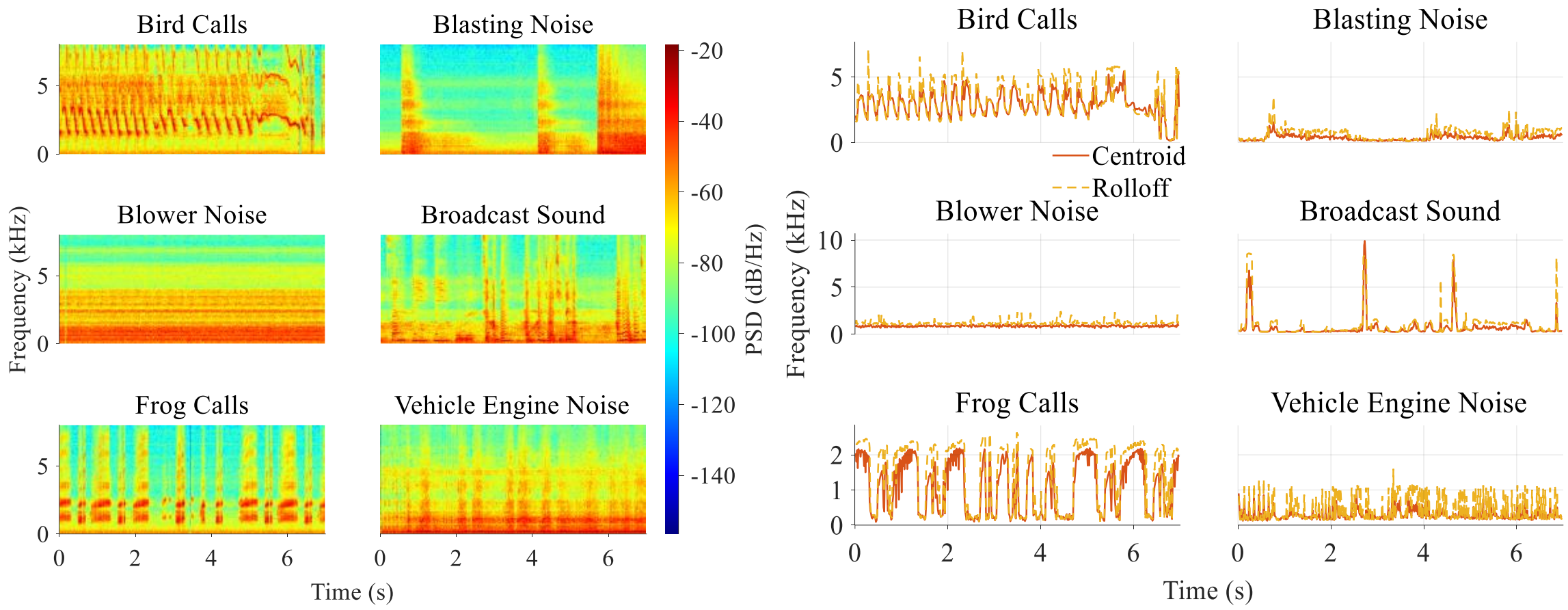

(a) Time-frequency spectrograms (b) Temporal trajectories of spectral features

**Fig. 3.** Frequency-domain analysis of the six representative sounds.

The above analysis indicates that there are significant directional differences in the characteristics of environmental sound on the time and frequency axes: the time dimension mainly describes the dynamic evolution process of acoustic events, while the frequency dimension characterizes the energy distribution on different frequency bands. The conventional square two-dimensional convolutions use the same local receptive field to jointly model both dimensions. Although they can extract the local time-frequency features, they have difficulty independently modeling temporal dynamics and frequency-band structures according to their different physical meanings. To solve this limitation, a time-frequency convolution module (FTConv) in the front-end of FTANet is designed, which uses two parallel asymmetric one-dimensional convolution branches to extract features in the frequency and time directions, respectively. Then, an adaptive fusion attention (AFA) module is introduced to recalibrate the fused time-frequency features. The FTANet architecture is shown in Fig. 1, and the FTConv and AFA modules are the key components of FTANet.

### 2.2.1. Frequency-Time Convolution Module (FTConv)

The FTConv module is used to extract the time-domain and the frequency-domain features from environmental acoustic signals. Its structure is shown in Fig. 4.

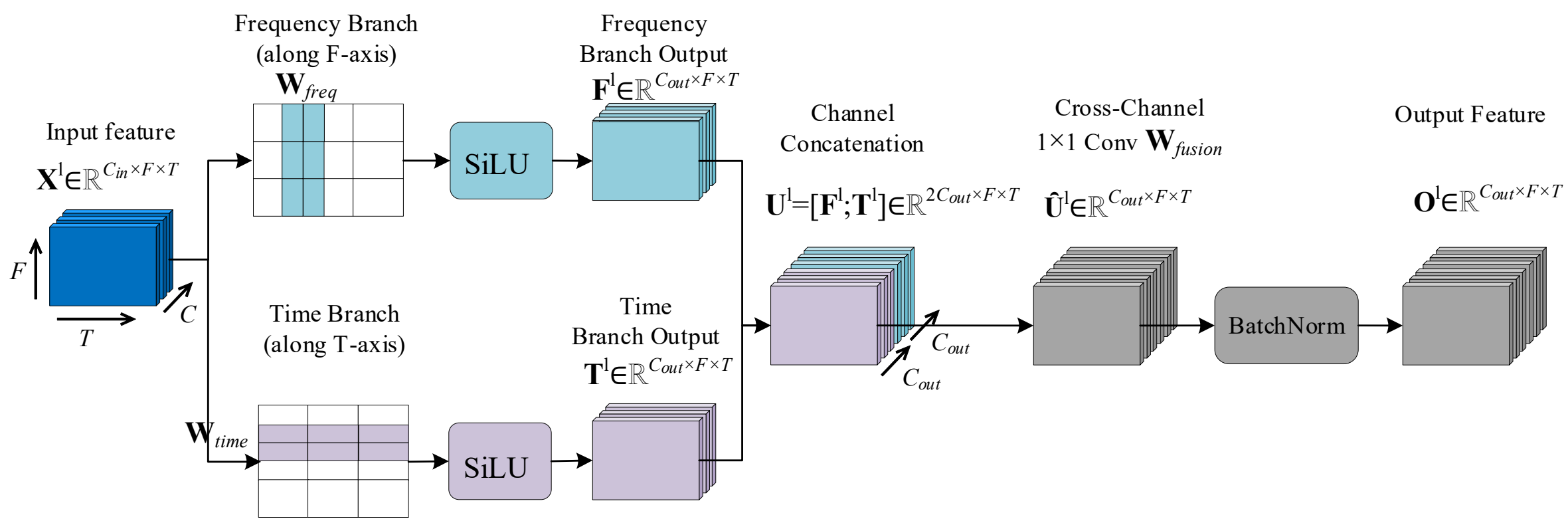

**Fig. 4.** Structure of the FTConv module.

As shown in Fig. 4, the forward outputs of the FTConv module are given by:

$$\begin{cases} \mathbf{F}^{l}_{c,f,t} = \mathrm{SiLu}\left(\sum_{k=1}^{C_{in}}\sum_{i=-1}^{1}\mathbf{W}^{3\times1}_{freq,c,k,i}\cdot\mathbf{X}^{l}_{k,f+i,t}\right) \\ \mathbf{T}^{l}_{c,f,t} = \mathrm{SiLu}\left(\sum_{k=1}^{C_{in}}\sum_{j=-1}^{1}\mathbf{W}^{1\times3}_{time,c,k,j}\cdot\mathbf{X}^{l}_{k,f,t+j}\right) \end{cases} \tag{1}$$

where $\mathbf{X}^l$ denotes the input feature of the $l$-th layer. The frequency-branch kernel $\mathbf{W}_{freq}\in\mathbb{R}^{Cout\times Cin\times 3\times 1}$ and the time-branch convolution kernel $\mathbf{W}_{time}\in\mathbb{R}^{Cout\times Cin\times 1\times 3}$ are used in the frequency and time branches, respectively. The intermediate joint representation obtained through concatenation is given by $\mathbf{U}^l=[\mathbf{F}^l;\mathbf{T}^l]\in\mathbb{R}^{2Cout\times F\times T}$. It is passed through a 1 × 1 convolution to perform cross-channel linear interactions, followed by batch normalization and SiLU activation to achieve feature fusion and scale alignment:

$$\mathbf{O}^l = SiLU\left(\frac{\hat{\mathbf{U}}^l - \mathrm{E}[\hat{\mathbf{U}}^l]}{\sqrt{Var[\hat{\mathbf{U}}^l]+\epsilon}}\cdot\gamma+\beta\right) \tag{2}$$

where $\hat{\mathbf{U}}^l=\mathbf{W}^{1\times 1}_{\text{fusion}}*\mathbf{U}^l$ denotes the feature map obtained by applying the cross-channel interaction convolution kernel $\mathbf{W}^{1\times 1}_{\text{fusion}}$ to $\mathrm{E}[\hat{\mathbf{U}}^l]$ and $\mathrm{Var}[\hat{\mathbf{U}}^l]$ denote the mean and variance, respectively, of this feature map over the current mini-batch. The parameters $\gamma$ and $\beta$ are the learnable scale and shift parameters of the batch normalization layer, respectively. By using this method above, the independently extracted time features and frequency features are mapped into an unified feature space, as the input for the subsequent adaptive fusion attention module.

### 2.2.2. Adaptive Fusion Attention Mechanism

After dual-branch feature extraction by FTConv and feature fusion through the 1 × 1 convolution, the time and frequency features directions undergo preliminary cross-channel interaction. However, different channels have different contributions to the discrimination of the sound categories. If the same weight is assigned to all channels, it may result in more redundant responses. Therefore, an AFA module is added after feature fusion. Adaptively learn the importance of each channel based on input features, and recalibrate the fused time-frequency features. The structure of AFA is shown in Fig. 5.

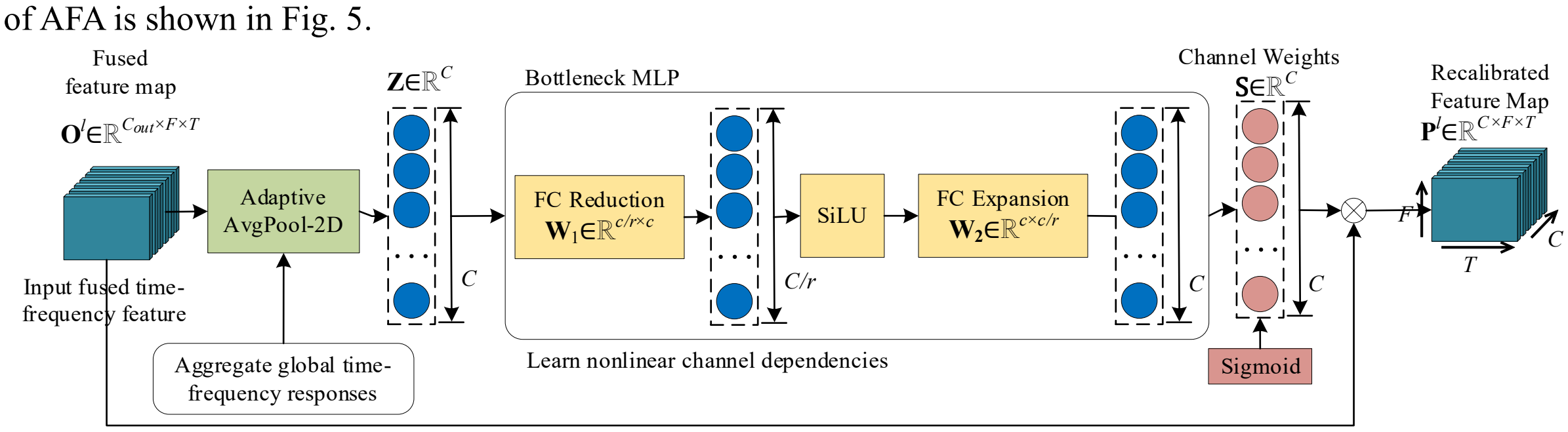

**Fig. 5.** Structure of the AFA module.

Let $\mathbf{O}^l\in\mathbb{R}^{C\times F\times T}$ denote the fused feature obtained from FTConv. In AFA, the global average pooling is first performed in both frequency and time dimensions, and then the two-dimensional time-frequency response is compressed into a one-dimensional vector $\mathbf{Z}\in\mathbb{R}^c$. The output of the $c$-th channel is given by:

$$\mathbf{Z}_c = \frac{1}{F\times T}\sum_{f=1}^{F}\sum_{t=1}^{T}\mathbf{O}^l_{c,f,t} \tag{3}$$

where $F$ and $T$ denote the frequency and time dimensions of the feature map, respectively. By using this method, the global time-frequency response of each channel is aggregated into a scalar to obtain channel information that contains the global signal.

Subsequently, AFA employs a two-layer fully connected structure to model nonlinear dependencies among channels. Firstly, the reduction layer $\mathbf{W}_1$ compresses the channel dimension according to the reduction ratio $r$, producing a low-dimensional bottleneck representation. After SiLU nonlinear activation, the restoration layer $\mathbf{W}_2$ maps the representation back to the original channel dimension. Finally, a normalized weight for each channel is generated by using the Sigmoid function :

$$\mathbf{S} = \delta\left(\mathbf{W}_2\cdot\mathrm{SiLU}(\mathbf{W}_1\cdot\mathbf{z})\right) \tag{4}$$

where $\mathbf{W}_1$ and $\mathbf{W}_2$ denote the learnable parameters used for channel reduction and restoration, respectively, and $\delta(\cdot)$ is the Sigmoid activation function. By using the bottleneck structure, the nonlinear relationships between different channels while reducing parameter overhead, and constrains the weights of each channel within the range of [0,1].

After obtaining the weight vector $\mathbf{S}$, it is applied back to the original fused feature $\hat{\mathbf{U}}^l$ to obtain the recalibrated feature $\mathbf{P}^l$, i.e.,

$$\mathbf{P}^{l}_{c,f,t} = \mathbf{S}_c \cdot \hat{\mathbf{U}}^{l}_{c,f,t}, \quad \forall c \in \{1, 2, \ldots, 2C_{out}\} \tag{5}$$

Dynamically adjust the response intensity of different channels based on input samples, so that the network pays more attention to channels with higher discriminative power for the current sound category, while reducing relatively redundant feature responses.

### 2.2.3. Multi-Stage Feature Modeling and Classification Output

After recalibrating the channel characteristics by using AFA, FTANet models the features through residual learning and max pooling. A progressive three-stage structure is employed to gradually extract time-frequency representations at different scales. Each stage consists of FTConv, AFA, a residual module, and max pooling. The output of the preceding stage serves as the input to the subsequent stage, allowing the features to acquire higher-level acoustic representations while their time-frequency resolution is progressively reduced. After three stages of feature modeling, the final features are sequentially passed through global average pooling, Dropout, and a fully connected layer to obtain the class scores. The multi-stage feature modeling and classification output of FTANet is shown in Fig. 6.

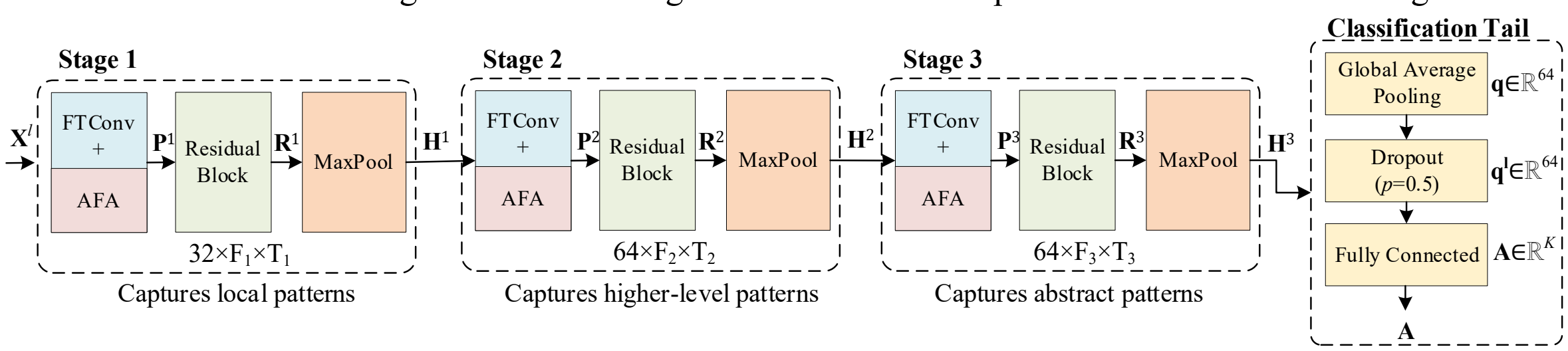

**Fig. 6.** Structure of multi-stage feature modeling and classification output in FTANet

Taking the AFA output $\mathbf{P}^l$ as its input, the two 3 × 3 convolutions are firstly used to learn an incremental representation of the input features in the residual module. This representation is then added to the original input through an identity mapping. The output can be expressed as:

$$\mathbf{R}^{l} = SiLU(\mathbf{P}^{l} + D^{l}_{res}(\mathbf{P}^{l})) \tag{6}$$

where $D^{l}_{res}$ denotes the residual mapping in the $l$-th stage, which is composed of two 3 × 3 convolutional layers together with batch normalization and nonlinear activation, and $\mathbf{R}^l$ denotes the output of the residual module. While further extracting local time-frequency patterns, this structure preserves the original information in the AFA-recalibrated features through the identity connection. This helps reduce information loss during multi-stage feature transformations and improves network optimization.

A 2 × 2 max-pooling is subsequently applied to down sample the residual features along both the frequency and time dimensions:

$$\mathbf{H}^{l} = \text{MaxPool}_{2\times 2}(\mathbf{R}^{l}) \tag{7}$$

In Eq. (7), the local responses are preserved while reducing the feature-map resolution and computational complexity in subsequent stages. It also allows subsequent convolutions to cover a larger effective time-frequency range.

Based on the above structure, FTANet employs three stages of feature modeling. The three stages use the same "FTConv-AFA-Residual-MaxPool" processing pattern but with different resolutions and numbers of channels. In the stage 1, a three-channel time-frequency feature map is taken as inputs. The time and frequency branches of FTConv each generates 16 channels, which are concatenated to form a 32-channel feature map. After AFA, residual learning, and 2 × 2 max pooling, the first-stage output is obtained as $\mathbf{H}^1 \in \mathbb{R}^{32\times F/2\times T/2}$. The stage 2 takes $\mathbf{H}^1$ as input and increases the number of output channels of each time-frequency branch to 32. The outputs are concatenated to form a 64-channel feature map. The same feature recalibration, residual learning, and down sampling are then applied to obtain $\mathbf{H}^2 \in \mathbb{R}^{64\times F/4\times T/4}$, thereby enhancing feature representation capability. In the stage 3, $\mathbf{H}^2$ is further used as input and time-frequency feature modeling is continued at a lower time-frequency resolution. Meanwhile, the number of output channels is maintained at 64 to avoid the additional parameter overhead caused by a further increase in network width. The final backbone output is obtained as $\mathbf{H}^3 \in \mathbb{R}^{64\times F/8\times T/8}$.

To convert the two-dimensional time-frequency features into a compact feature representation suitable for classification, the global average pooling (GAP) is first applied to aggregate $\mathbf{H}^3$ along the frequency and time dimensions. The pooling result for the $c$-th channel is expressed as:

$$\mathbf{q}_c = \frac{1}{F_3 \times T_3} \sum_{f=1}^{F_3} \sum_{t=1}^{T_3} \mathbf{H}^3_{c,f,t}, c = 1, 2, \ldots, C_3 \tag{8}$$

This produces a global channel feature map of size $C_3 \times 1 \times 1$. After flattening, it forms a one-dimensional acoustic feature vector $\mathbf{q}=[\mathbf{q}_1,\mathbf{q}_2,\ldots \mathbf{q}_{C3}]^{\mathrm{T}} \in \mathbb{R}^{C3}$. During training, Dropout regularization with a drop probability of 0.5 is applied to **q** to reduce the risk of model overfitting. The feature vector obtained after Dropout is denoted by $\mathbf{q}^l$. Finally, the regularized feature vector $\mathbf{q}^l$ is fed into the fully connected classification layer and mapped to the target class space, producing the class-score vector **A**:

$$\mathbf{A} = \mathbf{W}_{cls}\mathbf{q}^l + \mathbf{b}_{cls} \tag{9}$$

where $\mathbf{W}_{cls} \in \mathbb{R}^{K \times C3}$ and $\mathbf{b}_{cls} \in \mathbb{R}^K$ denote the weight matrix and bias vector of the fully connected classification layer, respectively. $K$ denotes the number of classes in the current classification task, and $\mathbf{A}=[a_1,a_2,\ldots,a_k]^{\mathrm{T}}$ denotes the class-score vector output by the network.

## 2.3. Two-Stage Hierarchical Classification Architecture

To alleviate cross category confusion caused by directly using a single-stage classifier, , a two-stage hierarchical classification method is constructed in this paper. In this method, the FTANet's internal network architecture is not changed. Instead, it organizes multiple independent FTANet classification models based on the hierarchical relationships between environmental sound labels, and divides the classification process into two stages: the coarse-grained category discrimination and the within-category fine-grained classification. The two-stage hierarchical classification architecture is shown in Fig. 1(the right component of Fig.1)

Assuming the set of coarse-grained categories is $\mathbf{M}=\{m_1, m_2, \ldots, m_5\}$, i.e., the industrial sounds, the construction sounds, the transportation sounds, the social activity sounds, and the natural sounds. Each coarse-grained category $m_k$ has a fine-grained label subset $Y_k \subset Y_{all}$. These subsets satisfy the mutual-exclusivity and completeness constraints, i.e.,

$$\begin{cases} \bigcup_{k=1}^{5} Y_k = \{y_1, y_2, \ldots, y_{25}\} \\ Y_i \cap Y_j = \varnothing \quad (\forall i \neq j) \end{cases} \tag{10}$$

In the first stage, the FTANet classification model is trained, which consists of five coarse-grained categories. Let this model be denoted by $V^{(1)}(;\mathbf{\Theta}^{(1)})$, where $\mathbf{\Theta}^{(1)}$ denotes all learnable parameters of the first-stage FTANet. Assuming the input feature is $\mathbf{X}^l$, the first-stage model outputs a five-dimensional class-score vector, i.e.,

$$\mathbf{A}^{(1)} = V^{(1)}(\mathbf{X};\mathbf{\Theta}^{(1)}) = [a_1^{(1)}, a_2^{(2)}, \ldots, a_5^{(5)}]^T \in \mathbb{R}^5 \tag{11}$$

Where $\mathbf{A}^{(1)}$ is the class-score vector produced by the FTANet model after three stages of feature modeling, global average pooling, Dropout, and the fully connected classification layer. The softmax function is then used to obtain the probability that the input sample belongs to the five coarse-grained categories, i.e.,

$$p_k^{(1)} = P(m_k | \ \mathbf{X}) = \frac{\exp(a_k^{(1)})}{\sum_{i=1}^{5} \exp(a_i^{(1)})}, k = 1, 2, \ldots, 5. \tag{12}$$

The coarse-grained category with the highest first-stage classification probability is then selected. Its category index is expressed as $k^* = \operatorname{argmax}_{k \in \{1,2,\ldots,5\}} p_k^{(1)}$, where $k^*$ determines the fine-grained subclassifier invoked in the second stage. During inference, only the second-stage FTANet corresponding to the predicted coarse-grained category $m_{k^*}$ is invoked. Consequently, the subsequent classification is restricted to the corresponding within-category fine-grained label space.

Assuming the five second-stage sub-classifiers are denoted by $V_1^{(2)}$, $V_2^{(2)}, \ldots, V_5^{(2)}$. The $k$-th second-stage classifier is trained only by the samples belonging to the coarse-grained category $m_k$, and its classification label space is $Y_k$. Once $k^*$ has been predicted by the first stage, the original preprocessed input feature **X** is transmitted to the corresponding second-stage FTANet, i.e.,

$$\mathbf{A}_{k^*}^{(2)} = V_{k^*}^{(2)}(\mathbf{X};\mathbf{\Theta}_{k^*}^{(2)}), \mathbf{A}_{k^*}^{(2)} \in \mathbb{R}^{K_{k^*}} \tag{13}$$

where $\mathbf{\Theta}_{k^*}^{(2)}$ denotes the learnable parameters of the $k^*$-th second-stage FTANet, and $K_{k^*}=|Y_{k^*}|$ is the number of fine-grained classes contained in the selected coarse-grained category. For any $y \in Y_{k^*}$, the second-stage classification probability is expressed as:

$$p_y^{(2)} = P(y|\ \mathbf{X}, k^*) = \frac{\exp(a_{k^*,y}^{(2)})}{\sum_{j=Y_{k^*}}^{5} \exp(a_{k^*,j}^{(2)})}. \quad (14)$$

The final fine-grained classification result is determined by the class with the highest probability from the selected second-stage classifier: $\hat{y} = \arg\max_{y \in Y_k} p_y^{(2)}$.

Through the above hierarchical classification, the original global fine-grained classification task is decomposed into "first level coarse-grained category discrimination+second level intra class fine-grained recognition". The first-stage FTANet establishes decision boundaries among sound-source categories, whereas the second-stage FTANet further performs fine-grained discrimination among sounds that are easily confused within the same coarse-grained category, which reduces the category search space. Meanwhile, during inference, only the first-stage model and one second-stage model corresponding to the first-stage prediction are invoked, rather than all second-stage sub-classifiers working simultaneously.

## 3. Experimental Studies

### 3.1. Data Preprocessing and Training Strategy

The experiments are tested by using a self-constructed dataset of complex environmental sounds and a public dataset. The coarse-grained categories of the self-constructed dataset are defined with reference to the classification of noise sources in the *Law of the People's Republic of China on Noise Pollution Prevention and Control*, including the industrial noises, the construction noises, the transportation noises, and the social activity noises. Considering that real urban acoustic environments also contain a large number of natural background sounds, a natural sound category is added. A hierarchical classification system is constructed, whichconsists of five coarse-grained categories and their corresponding 25 fine-grained sound classes, and they are shown in Table 1.

**Table 1.** Label taxonomy of the self-constructed dataset.

| Category ID | Noise Category | Subclass ID | Sound Class |
|---|---|---|---|
| C1 | Industrial Noise | S01 | Machining Noise |
| C1 | Industrial Noise | S02 | Rotating Machinery Noise |
| C1 | Industrial Noise | S03 | Air Compressor Noise |
| C1 | Industrial Noise | S04 | Ventilation Fan Noise |
| C1 | Industrial Noise | S05 | Blower Noise |
| C2 | Natural Sounds | S06 | Insect Sounds |
| C2 | Natural Sounds | S07 | Frog Calls |
| C2 | Natural Sounds | S08 | Cicada Calls |
| C2 | Natural Sounds | S09 | Rain |
| C2 | Natural Sounds | S10 | Thunder |
| C2 | Natural Sounds | S11 | Wind |
| C2 | Natural Sounds | S12 | Bird Calls |
| C3 | Social Noise | S13 | Conversation |
| C3 | Social Noise | S14 | Broadcast Sound |
| C3 | Social Noise | S15 | Amplified Music/Sound |
| C4 | Construction Noise | S16 | Material-Handling Machinery Noise |
| C4 | Construction Noise | S17 | Mechanical Impact Noise |
| C4 | Construction Noise | S18 | Blasting Noise |
| C4 | Construction Noise | S19 | Electric Hammer/Pile Driver Noise |
| C4 | Construction Noise | S20 | Crane Operation Noise |
| C5 | Traffic Noise | S21 | Vehicle Braking Noise |
| C5 | Traffic Noise | S22 | Vehicle Engine Noise |
| C5 | Traffic Noise | S23 | Vehicle Horn |
| C5 | Traffic Noise | S24 | Train Rumble |
| C5 | Traffic Noise | S25 | Aircraft Roar |

The distribution of 25 categories sounds in Fig. 7.

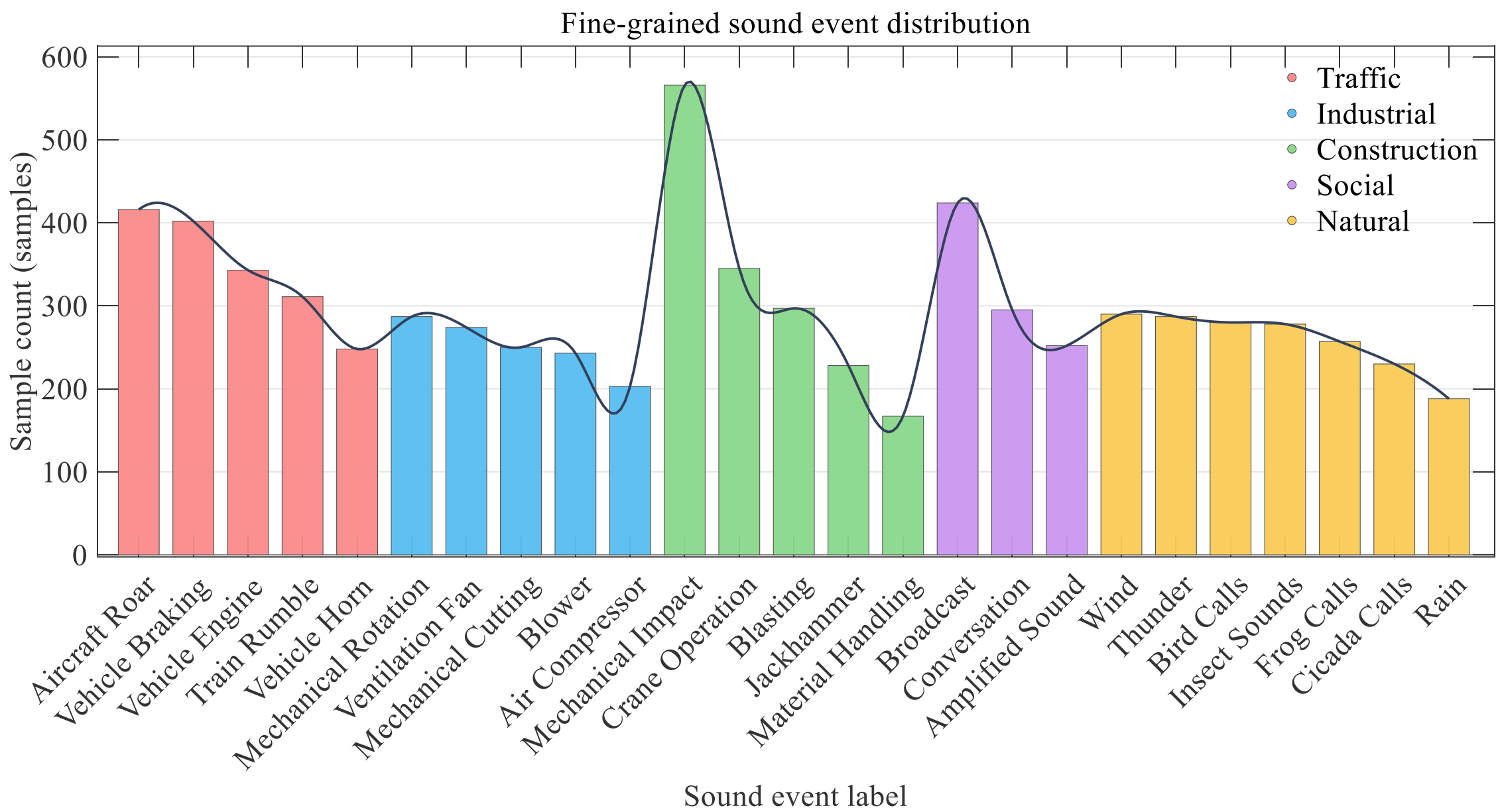

**Fig. 7.** Distribution of 25 categories sounds.

As shown in Fig. 7, the self-constructed dataset is relatively imbalance, i.e., some high-frequency environmental sounds have a larger number of samples, while some occasional or easily affected sound samples by background noise have a relatively smaller number. . The self-constructed dataset contains a total of 9,029 sounds samples. In the experiments, the samples in each fine-grained class are randomly divided into training sets, validation sets, and testing sets at a ratio of 6:2:2.

Environmental sound signal preprocessing primarily consists of signal resampling, Fbank feature extraction, and feature construction. Firstly, all audio recordings are converted to mono and resampled to $f_s$=16000 Hz to unify the sampling rate differences generated by different acquisition devices. Audio recordings with sampling rates different from 16 kHz were resampled using the sinc-interpolation resampling method implemented in Torchaudio, which incorporates anti-aliasing filtering. Mel filter-bank (Fbank) features are then extracted. Compared with Mel-frequency cepstral coefficients (MFCCs), the Fbank features preserve local correlations between frequency bands. The number of Mel filters is set to $M$=80, and the analysis window length is set to 25 ms, corresponding to $N_L = [0.025 \times f_s] = 400$ samples per frame.

To ensure that sound signals of different lengths have a unified time dimension before being input into the network, the frame shift $N_S$ is adaptively determined according to the target number of frames $T$ and the fixed window length $N_L$, i.e.,

$$N_S = \Phi(L, N_L, T) = \max\left(\left\lfloor \frac{L - N_L}{T - 1} \right\rfloor, 1\right) \tag{15}$$

where $L$ is the number of sound samples. Based on $N_L$ and $N_S$, the two-dimensional Fbank time-frequency characteristics of the sound signals are obtained by using the Short Time Fourier Transform (STFT) and Mel filter bank. On this basis, a three-channel input feature is further constructed to enhance the representation of the dynamic variations of sounds. Assuming the two-dimensional Fbank feature is denoted by $\mathbf{S} \in \mathbb{R}^{M \times T}$, the first-order differences and the second-order differences are calculated, and the two-dimensional time-frequency representation is extended to a three-channel tensor by using the mapping $\mathcal{J}: \mathbb{R}^{M \times T} \rightarrow \mathbb{R}^{3 \times M \times T}$. The three-channel tensor is defined as:

$$\begin{cases} \mathbf{X}_{1,f,t} = \mathbf{S}_{f,t} \\ \mathbf{X}_{2,f,t} = \nabla_t \mathbf{S}_{f,t} = \mathbf{S}_{f,t} - \mathbf{S}_{f,t-1} \\ \mathbf{X}_{3,f,t} = \nabla_f \mathbf{S}_{f,t} = \mathbf{S}_{f,t} - \mathbf{S}_{f-1,t} \end{cases} \tag{16}$$

The final input feature tensor is denoted by $\mathbf{X}^l \in \mathbb{R}^{3 \times M \times T}$, where the three channels contain the static Fbank feature, the time-direction first-order difference, and the frequency-direction first-order difference, respectively. The tensor serves as the standard input to FTANet.

A stage-wise independent training strategy is used, and the first-stage coarse classifier and the second-stage subclassifiers are separately trained. The first level classifier is used for discriminating 5 coarse-grained categories, while the second level sub classifier completes fine-grained sound classification. The Adam optimizer is used for all models, the initial learning rate is $5 \times 10^{-4}$, the weight decay is $1 \times 10^{-4}$, and the batch size is 32. The each model is trained for 15 epochs. All experiments are conducted by using the PyTorch framework on a single NVIDIA GeForce RTX 4090 GPU.

### 3.2. Performance Comparison with Conventional Models

To evaluate FTANet's recognition performance and model complexity for environmental sound classification, several typical networks are selected as comparison models: the conventional heavyweight convolutional networks ESResNet [27] and VGG16; the Transformer-based Audio Spectrogram Transformer (AST) [28]; and the lightweight networks 3 × 3 CNN, SSRP-CNN [29], and ESC-NAS [30]. The experiments are conducted on the self-constructed environmental sound dataset and the public UrbanSound8K dataset [31]. The results are written in Tables 2 and 3.

To evaluate the classification performances, the accuracy, the precision, the recall, and F1-score are adopted, i.e.,

$$\begin{cases} Precision_i = \dfrac{TP_i}{TP_i + FP_i} \\ Recall_i = \dfrac{TP_i}{TP_i + FN_i} \\ F1_i = \dfrac{2 Precision_i \times Recall_i}{Precision_i + Recall_i} \end{cases} \tag{17}$$

where $TP_i$, $FP_i$, and $FN_i$ denote the numbers of the true-positive samples, the false-positive samples, and the false-negative samples for the $i$-th category, respectively. The overall accuracy is calculated as

$$Accuracy = \frac{\sum_{i=1}^{C} TP_i}{N} \tag{18}$$

where $C$ is the number of categories and $N$ is the total number of test samples. To evaluate the performance across all categories, the macro-averaged F1-score is calculated as

$$Macro-F1 = \frac{1}{C}\sum_{i=1}^{C} F1_i \tag{19}$$

**Table 2.** Performance comparison on the self-constructed dataset with five coarse-grained categories.

| Method | **Category ID** | Precision | Recall | F1-score | Overall Accuracy | Macro-F1 | #Params(M) | Flops(G) |
|---|---|---|---|---|---|---|---|---|
| **FTANet** | C1 | 0.9604 | 0.9209 | 0.9402 | **0.9545** | **0.9523** | **0.2004** | **0.3011** |
| | C2 | 0.9539 | 1.0000 | 0.9764 | | | | |
| | C3 | 0.9498 | 0.9342 | 0.9419 | | | | |
| | C4 | 0.9507 | 0.9578 | 0.9543 | | | | |
| | C5 | 0.9574 | 0.9397 | 0.9485 | | | | |
| CNN | C1 | 0.8988 | 0.9343 | 0.9162 | 0.9300 | 0.9369 | 0.2232 | 0.3179 |
| | C2 | 0.9809 | 0.9879 | 0.9844 | | | | |
| | C3 | 0.8840 | 0.9805 | 0.9298 | | | | |
| | C4 | 0.9452 | 0.8772 | 0.9099 | | | | |
| | C5 | 0.9593 | 0.9298 | 0.9443 | | | | |
| VGG16 | C1 | 0.8891 | 0.9264 | 0.9074 | 0.9076 | 0.9173 | 134.2810 | 2.0771 |
| | C2 | 0.9894 | 0.9253 | 0.9563 | | | | |
| | C3 | 0.8689 | 0.9342 | 0.9003 | | | | |
| | C4 | 0.9294 | 0.8741 | 0.9009 | | | | |
| | C5 | 0.8996 | 0.9455 | 0.9219 | | | | |
| SSRP-CNN | C1 | 0.8818 | 0.9209 | 0.9009 | 0.9470 | 0.9436 | 0.3899 | 0.1421 |
| | C2 | 0.9912 | 0.9956 | 0.9934 | | | | |
| | C3 | 0.8996 | 0.9959 | 0.9453 | | | | |
| | C4 | 0.9381 | 0.9032 | 0.9204 | | | | |
| | C5 | 0.9901 | 0.9281 | 0.9581 | | | | |
| ESResNet | C1 | 0.8237 | 0.9462 | 0.8807 | 0.9334 | 0.9306 | 11.1791 | 1.8235 |
| | C2 | 0.9782 | 0.9846 | 0.9814 | | | | |
| | C3 | 0.9192 | 0.9835 | 0.9503 | | | | |
| | C4 | 0.9535 | 0.8139 | 0.8782 | | | | |
| | C5 | 0.9716 | 0.9536 | 0.9625 | | | | |
| | C1 | 0.8854 | 0.9051 | 0.8951 | | | | |
| | C2 | 0.9145 | 0.8703 | 0.8919 | | | | |
| ESC-NAS | C3 | 0.9426 | 0.8107 | 0.8717 | 0.8853 | 0.8855 | 0.0482 | 0.0139 |
| | C4 | 0.9335 | 0.9057 | 0.9194 | | | | |
| | C5 | 0.7967 | 0.9095 | 0.8494 | | | | |
| | C1 | 0.9614 | 0.9462 | 0.9537 | | | | |
| | C2 | 0.9891 | 1.0000 | 0.9945 | | | | |
| AST | C3 | 0.9275 | 1.0000 | 0.9624 | 0.9648 | 0.9634 | **85.2588** | **51.2949** |
| | C4 | 0.9569 | 0.9355 | 0.9460 | | | | |
| | C5 | 0.9715 | 0.9490 | 0.9601 | | | | |

As shown in Table 2, based on the self-constructed environmental sound dataset with five coarse-grained categories, FTANet the accuracy of FTANet is 95.45% and the Macro-F1 score is 95.23%, which is second only to AST. In AST, the accuracy is 96.48% and the Macro-F1 is 96.34%, which is exceeding FTANet by 1.03% and 1.11%, respectively. However, AST has 85.2588 M parameters and 51.2949 G FLOPs, whereas FTANet has only 0.2004 M parameters and 0.3011 G FLOPs, which greatly reduces the computational complexity.

The advantages of FTANet are also evident in comparisons with the convolutional and lightweight models. Compared with the 3 × 3 CNN, FTANet increases accuracy from 93.00% to 95.45%, while reducing the parameter and FLOPs by approximately 10.22% and 5.28%, respectively. FTANet also outperforms SSRP-CNN by 0.75% in accuracy while using approximately 48.60% fewer parameters, although SSRP-CNN has fewer FLOPs. Compared with the heavyweight VGG16 and ESResNet models, the FTANet improves the accuracy by 4.69% and 2.11%, respectively, while requiring substantially fewer parameters and FLOPs. Although ESC-NAS has the fewest parameters and FLOPs, its accuracy is 88.53%, which is lower than that of FTANet. In a word, the FTANet provides a favorable balance between recognition performance and model complexity, which demonstrates its advantages for deployment in resource-constrained environmental sound classification systems.

To further validate the generalization performance of FTANet on a public dataset, the experiments are tested by following the 10-fold data partition of UrbanSound8K. The experimental results are shown in Table 3.

**Table 3.** Experimental results by using the UrbanSound8K dataset

| | **FTANet** | | CNN | | VGG16 | | SSRP-CNN | | ESResNet | | ESC-NAS | | AST | |
|---|---|---|---|---|---|---|---|---|---|---|---|---|---|---|
| | Accuracy | F1-score | Accuracy | F1-score | Accuracy | F1-score | Accuracy | F1-score | Accuracy | F1-score | Accuracy | F1-score | Accuracy | F1-score |
| Fold1 | 0.7446 | 0.7686 | 0.7446 | 0.7781 | 0.6861 | 0.7269 | 0.7560 | 0.7684 | 0.7331 | 0.7574 | 0.6231 | 0.6370 | 0.8797 | 0.8974 |
| Fold2 | 0.8176 | 0.8206 | 0.8041 | 0.8108 | 0.6993 | 0.7070 | 0.7354 | 0.7562 | 0.7162 | 0.7436 | 0.6374 | 0.6637 | 0.8986 | 0.9099 |
| Fold3 | 0.7459 | 0.7733 | 0.7211 | 0.7329 | 0.6443 | 0.6818 | 0.6649 | 0.6985 | 0.6886 | 0.7023 | 0.5405 | 0.5608 | 0.7741 | 0.8012 |
| Fold4 | 0.7859 | 0.7772 | 0.8192 | 0.8158 | 0.6909 | 0.6789 | 0.7798 | 0.7896 | 0.7515 | 0.7461 | 0.6606 | 0.6636 | 0.8889 | 0.8857 |
| Fold5 | 0.8291 | 0.8311 | 0.7788 | 0.7703 | 0.8024 | 0.8128 | 0.8451 | 0.8462 | 0.8462 | 0.8524 | 0.6688 | 0.6700 | 0.9327 | 0.9383 |
| Fold6 | 0.8214 | 0.8305 | 0.8688 | 0.8802 | 0.7789 | 0.7921 | 0.7825 | 0.8006 | 0.8117 | 0.8271 | 0.6877 | 0.7050 | 0.8979 | 0.9076 |
| Fold7 | 0.7780 | 0.7863 | 0.8103 | 0.8186 | 0.7136 | 0.7085 | 0.7613 | 0.7745 | 0.7494 | 0.7590 | 0.6038 | 0.6191 | 0.8687 | 0.8513 |
| Fold8 | 0.7618 | 0.7891 | 0.6960 | 0.7266 | 0.6675 | 0.6756 | 0.6824 | 0.6964 | 0.7246 | 0.7378 | 0.6811 | 0.6875 | 0.8697 | 0.8817 |
| Fold9 | 0.8174 | 0.8307 | 0.8591 | 0.8704 | 0.7733 | 0.7840 | 0.8125 | 0.8132 | 0.7904 | 0.8094 | 0.7316 | 0.7483 | 0.9020 | 0.9148 |
| Fold10 | 0.8220 | 0.8298 | 0.7885 | 0.7963 | 0.7694 | 0.7819 | 0.8124 | 0.8252 | 0.8041 | 0.8169 | 0.7013 | 0.7014 | 0.9140 | 0.9225 |
| Mean | **0.7924** | **0.8037** | 0.7891 | 0.8000 | 0.7226 | 0.7349 | 0.7632 | 0.7769 | 0.7616 | 0.7752 | 0.6536 | 0.6656 | 0.8826 | 0.8910 |
| **Δ**Mean | -- | -- | **0.0033** | **0.0037** | **0.0698** | **0.0688** | **0.0292** | **0.0268** | **0.0308** | **0.0285** | **0.1388** | **0.1381** | -0.0899 | -0.0888 |

As shown in Table 3, FTANet achieves an average accuracy of 79.24% and an average F1-score of 80.37% in the ten-fold evaluation on the UrbanSound8K dataset. As indicated by the **Δ**MeanFTANet results(i.e., the error between the mean of FTANet and the mean of other methods), FTANet outperforms the 3 × 3 CNN by 0.33% and 0.37% in the mean accuracy and the mean F1-score, respectively. Compared with VGG16, SSRP-CNN, ESResNet, and ESC-NAS, FTANet improves the mean accuracy by 6.98%, 2.92%, 3.08%, and 13.88%, respectively, while improving the mean F1-score by 6.88%, 2.68%, 2.85%, and 13.81%. Therefore, FTANet achieves the highest mean accuracy and F1-score among the compared convolution-based models on UrbanSound8K.

In AST, the mean accuracy and the F1-score are higher than those of FTANet by 8.99% and 8.88%, respectively. However, as shown in Table 2, AST has 85.2588 M parameters and 51.2949 G FLOPs, whereas FTANet has only 0.2004 M parameters and 0.3011 G FLOPs. Thus, FTANet provides competitive classification performance at a substantially lower model complexity. In summary, the FTANet achieves a favorable balance between classification performance, generalization capability, and computational complexity.

### 3.3. Comparison of Single-Stage and Two-Stage Classification Architectures

To verify the effectiveness of the two-stage hierarchical classification strategy for fine-grained environmental sound recognition, the FTANet is used as a common feature-extraction backbone to construct a single-stage classification model and a two-stage hierarchical classification model for comparison. The single-stage model directly maps the extracted acoustic features to all 25 fine-grained classes. By contrast, in the two-stage hierarchical classification model, a first-stage coarse classifier is firstly used for determining the broad category to which the sample belongs, and then the corresponding second-stage sub-classifier is selected to perform within-category fine-grained recognition. The experimental results are shown in Table 4.

**Table 4.** Experimental results of single-stage and two-stage classification on the 25-class dataset.

| **Coarse class** | Fine class | Single-stage | | | Two-stage | | | **ΔF1↑** |
|---|---|---|---|---|---|---|---|---|
| | | Precision | Recall | F1-score | Precision | Recall | F1-score | |
| C1 | S01 | 0.7500 | 0.4762 | 0.5825 | 0.8542 | 0.6508 | 0.7387 | **0.1562** |
| | S02 | 0.9333 | 0.9722 | 0.9524 | 0.9863 | 1.0000 | 0.9931 | **0.0407** |
| | S03 | 0.6286 | 0.8627 | 0.7273 | 0.9608 | 0.9608 | 0.9608 | **0.2335** |
| | S04 | 0.7667 | 1.0000 | 0.8679 | 0.7841 | 1.0000 | 0.8790 | **0.0111** |
| | S05 | 0.5962 | 0.5082 | 0.5487 | 0.6613 | 0.6721 | 0.6667 | **0.1180** |
| C2 | S06 | 0.6667 | 0.8857 | 0.7607 | 0.6897 | 0.8571 | 0.7643 | **0.0036** |
| | S07 | 0.7292 | 0.5385 | 0.6195 | 0.8533 | 0.9846 | 0.9143 | **0.2948** |
| | S08 | 0.7586 | 0.7586 | 0.7586 | 0.8627 | 0.7586 | 0.8073 | **0.0487** |
| | S09 | 1.0000 | 1.0000 | 1.0000 | 0.9592 | 1.0000 | 0.9792 | -0.0208 |
| | S10 | 0.9859 | 0.9722 | 0.9790 | 1.0000 | 0.9722 | 0.9859 | **0.0069** |
| | S11 | 0.9531 | 0.8356 | 0.8905 | 0.9467 | 0.9726 | 0.9595 | **0.069** |
| | S12 | 0.5641 | 0.3143 | 0.4037 | 0.6735 | 0.4714 | 0.5546 | **0.1509** |
| C3 | S13 | 0.9706 | 0.8919 | 0.9296 | 0.8795 | 0.9865 | 0.9299 | **0.0003** |
| | S14 | 0.8966 | 0.9811 | 0.9369 | 0.8455 | 0.9811 | 0.9083 | -0.0286 |
| | S15 | 0.9130 | 0.6667 | 0.7706 | 0.9344 | 0.9048 | 0.9194 | **0.1488** |
| C4 | S16 | 0.3333 | 0.0238 | 0.0444 | 0.5000 | 0.3333 | 0.4000 | **0.3556** |
| | S17 | 1.0000 | 1.0000 | 1.0000 | 0.9930 | 1.0000 | 0.9965 | -0.0035 |
| | S18 | 0.7234 | 0.4533 | 0.5574 | 0.7647 | 0.3467 | 0.4771 | -0.0803 |
| | S19 | 0.6667 | 0.7018 | 0.6838 | 0.6049 | 0.8596 | 0.7101 | **0.0263** |
| | S20 | 0.6148 | 0.9540 | 0.7477 | 0.8058 | 0.9540 | 0.8737 | **0.1260** |
| C5 | S21 | 0.6043 | 0.8317 | 0.7000 | 0.6977 | 0.8911 | 0.7826 | **0.0826** |
| | S22 | 0.2024 | 0.1977 | 0.2000 | 0.5940 | 0.9186 | 0.7215 | **0.5215** |
| | S23 | 0.8679 | 0.7419 | 0.8000 | 0.9149 | 0.6935 | 0.7890 | -0.0110 |
| | S24 | 1.0000 | 0.1410 | 0.2472 | 0.8333 | 0.7051 | 0.7639 | **0.5167** |
| | S25 | 0.3198 | 0.6058 | 0.4186 | 0.7692 | 0.2885 | 0.4196 | **0.0010** |
| **Overall Accuracy↑** | | | 0.7132 | | | **0.8149** | | |
| **Macro-F1↑** | | | 0.6851 | | | **0.7958** | | |
| Time(s) | | | 0.2170 | | | 1.2390 | | |

As shown in Table 4, in the single-stage classification model, the overall accuracy and the Macro-F1 score is respectively 71.32% and 68.51% across the 25 fine-grained environmental sound classes. After the two-stage hierarchical classification is adopted, the overall accuracy increases to 81.49% and the Macro-F1 score increases to 79.58%, which improves 10.17% and 11.07%, respectively. At the class level, the two-stage method improves the F1-score for 20 of the 25 fine-grained classes, indicating that the performance improvement is not limited to only a few categories. These results demonstrate the effectiveness of the two-stage hierarchical classification.

In terms of inference efficiency, the single-stage model requires only one forward classification pass, resulting in an inference time of 0.2170 s. The two-stage hierarchical classification model requires the first-stage coarse classification and second-stage fine-grained classification to be performed sequentially, increasing the inference time to 1.2390 s. Thus, although two-stage hierarchical classification improves recognition accuracy, it also requires additional inference overhead, which reflects a trade-off between classification performance and computational efficiency. To further analyze the differences in discriminative performance between the two classification approaches across different fine-grained classes, the confusion matrices are shown in Fig.8.

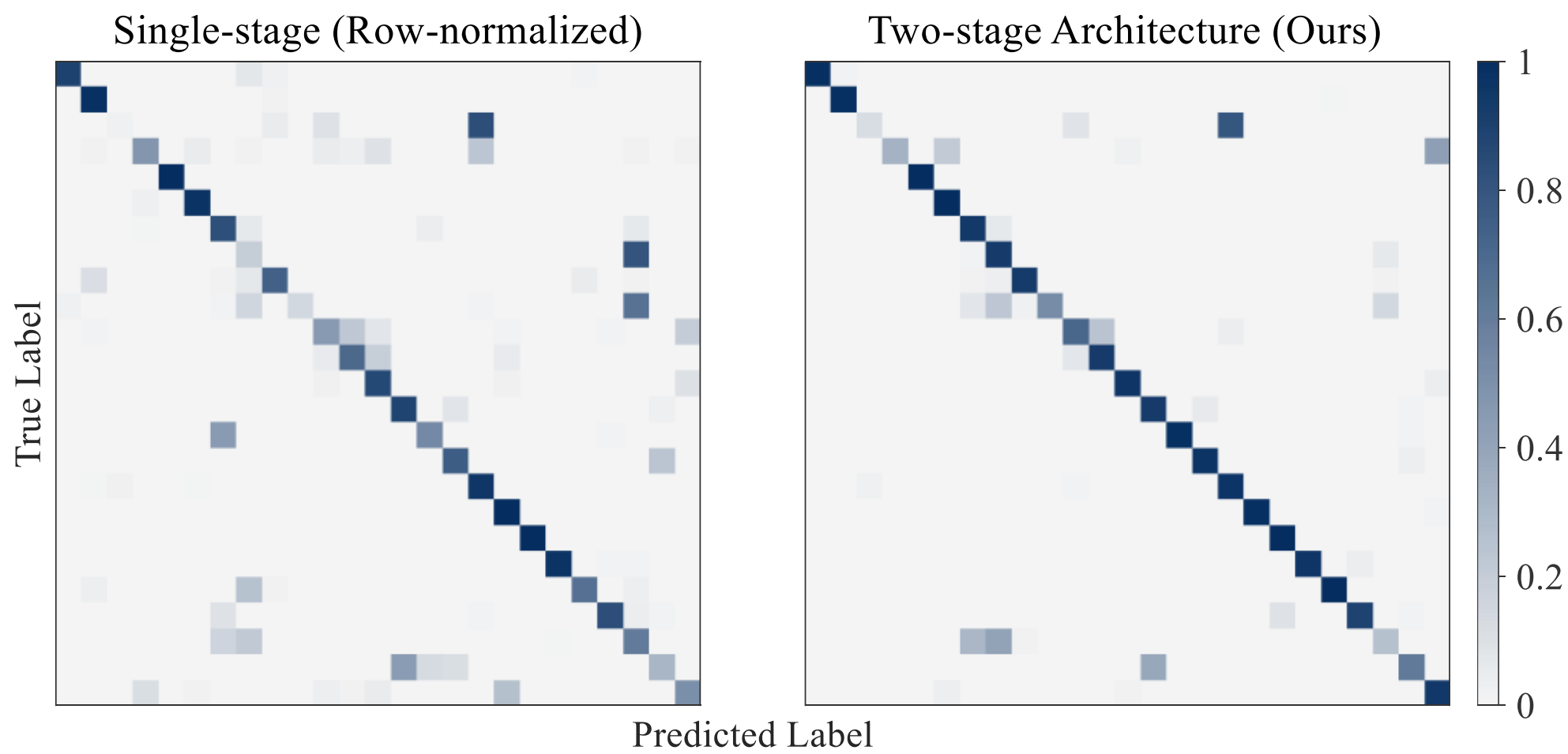

**Fig. 8.** Confusion matrices of the two methods across all fine-grained classes.

As shown in Fig. 8, the single-stage classification model exhibits relatively clear misclassification among several semantically similar classes. In the two-stage hierarchical classification model, the diagonal responses are more concentrated for most classes, and some intercategory confusion is alleviated. To further examine the feature distributions by using the two classification methods, t-distributed stochastic neighbor embedding (t-SNE) was used to perform dimensionality-reduction visualization of the high-dimensional acoustic features. The results are shown in Fig. 9.

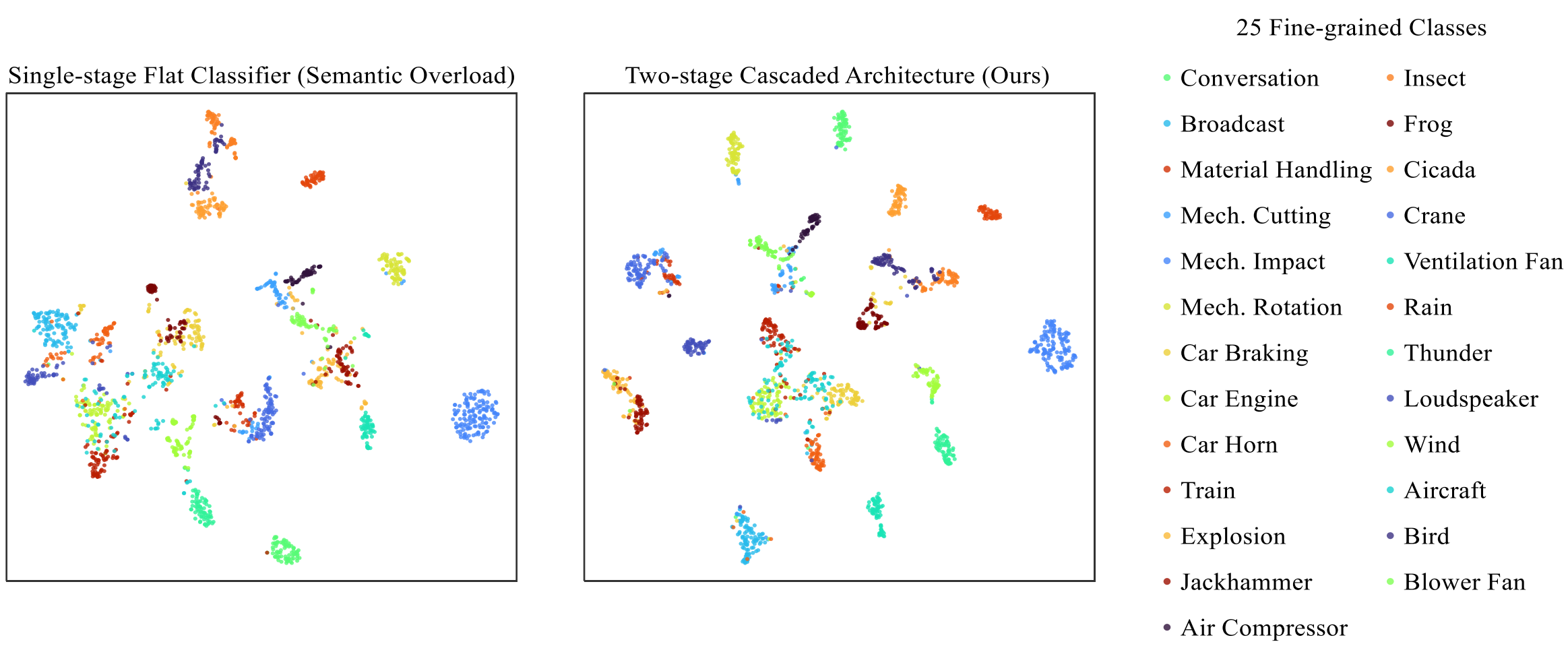

**Fig. 9.** t-SNE comparison of test-set feature distributions.

As shown in Fig. 9, the feature distributions of some fine-grained classes in the single-stage classification model exhibit relatively clear overlap. After two-stage hierarchical classification is adopted, the feature clusters of some different classes exhibit clearer clustering patterns and interclass boundaries. These visualization results are broadly consistent with the changes observed in the confusion matrices, which indicates that the hierarchical classification strategy can alleviate feature confusion among different classes.

### 3.4. Ablation Study of FTANet

An ablation study evaluates the contribution of each component of FTANet. Method A uses FTConv without 1 × 1 feature fusion or AFA; Method B uses FTConv and 1 × 1 feature fusion without AFA; Method C uses FTConv and AFA without 1 × 1 feature fusion; Method D uses standard 3 × 3 convolutions and AFA without 1 × 1 feature fusion; and Method E uses standard 3 × 3 convolutions without 1 × 1 feature fusion or AFA. The complete FTANet uses all three components. All other network structures and training settings remain identical across the configurations. Table 5 presents their classification accuracy, parameter counts, and FLOPs.

**Table 5.** Ablation study on the five coarse-grained classes.

| Method | Conv | | 1×1Fusion | AFA | Accuracy↑ | #Params(M) | Flops(G) |
|---|---|---|---|---|---|---|---|
| | 3 × 3 | FTConv | | | | | |
| **FTANet** | | √ | √ | √ | **0.9545** | **0.2004** | **0.3011** |
| A | | √ | | | 0.9359 | 0.1861 | 0.2847 |
| B | | √ | √ | | 0.9294 | 0.1958 | 0.3008 |
| C | | √ | | √ | 0.9394 | 0.1907 | 0.2850 |
| D | √ | | | √ | 0.9310 | 0.2281 | 0.3182 |
| E | √ | | | | 0.9300 | 0.2232 | 0.3179 |

A comparison between Methods E and A shows that it improves the model's accuracy from 93.00% to 93.59% by using the FTConv instead of the standard 3×3 convolution. Meanwhile, the parameter complexity decreases from 0.2232 M to 0.1861 M, which approximately reduces16.62%, and the FLOPs decrease from 0.3179 G to 0.2847 G, with a reduction of approximately 10.44%. These results demonstrate that it improves classification performance while reducing model complexity by using two asymmetric convolutional branches to separately model features in the time and frequency, thereby verifying the effectiveness of FTConv in modeling the time-frequency features of environmental sounds.

The AFA module improves the performances with different convolutional structures. Adding the AFA to the standard 3×3 convolutional baseline (Method D), the accuracy increases from 93.00% to 93.10%. When the AFA is Added to the FTConv (Method C), the accuracy increases from 93.59% to 93.94%. In both experiments, the AFA only added approximately 0.005 M additional parameters and 0.0003 G additional FLOPs. This indicates that the AFA can improve the channel feature representation capability of the network with relatively less additional computational overhead.

Further analysis of the 1×1 feature-fusion module shows that Method B ( i.e., adding feature fusion alone to the FTConv) reduces the accuracy from 93.59% to 92.94%, which indicates that the simple cross-channel linear fusion does not directly improve performance. By contrast, when the feature fusion method and the AFA are used together, the complete FTANet achieves an accuracy of 95.45%, which is higher than that of Method B by 2.51% and is higher than that of Method C by 1.51%. These results indicate that after the 1×1 convolution performs cross-channel interaction between the time and frequency branches, the AFA can further adjust the response strength of each channel according to the input features, thereby a more discriminative joint time-frequency feature representation can be obtained.

In summary, the complete FTANet has the highest accuracy of 95.45%, with a parameter count of 0.2004 M and FLOPs of 0.3011 G. Comparing with the standard 3×3 convolutional baseline, the Method E, the FTANet improves the accuracy by 2.45% while reducing the parameter count by approximately 10.22%. These results demonstrate that the combination of the FTConv, the feature fusion, and the AFA, can improve the performance of the environmental sound classification while reducing the computational complexity of the model.

## 4. Application

To verify the feasibility of the proposed hierarchical recognition method to the urban acoustic environment monitoring, the method is integrated into the Tianjin Online Acoustic Environment Monitoring Platform ( in Tianjin city, China) and tested in field. The application workflow of this method on the platform is shown in Fig. 10.

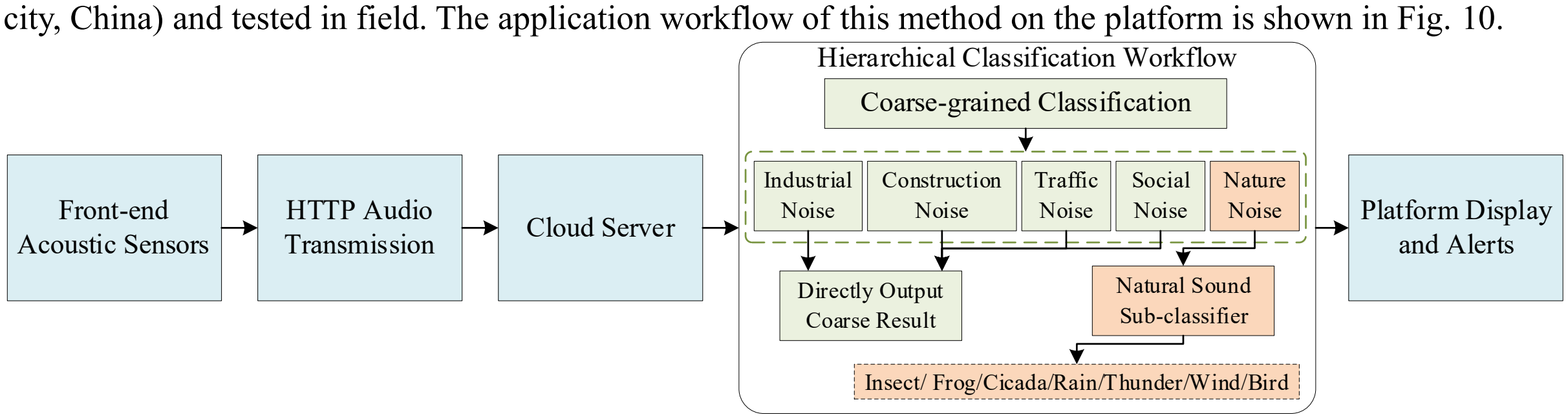

**Fig. 10.** Application workflow of the hierarchical classification method on the monitoring platform.

The front-end acoustic sensors are deployed in the urban street environments to continuously collect environmental sounds. The collected acoustic data are transmitted to a cloud server via HTTP. The server performs sound signal analysis, online recognition, and result generation, and then the recognition results are returned to the platform's display and early-warning modules.

Based on the existing platform classification system, the proposed hierarchical classification procedure is integrated into the online recognition pipeline. In the first-stage classification, the system discriminates the five coarse-grained categories, i.e., the industrial noise, the construction noise, the transportation noise, the social activity noise, and the natural sounds. The subsequent classification process is determined according to the first-stage prediction results. In this deployment, the system initially incorporated the second-stage sub-classifier for natural sounds to perform within-category fine-grained recognition. For the other four coarse-grained categories, the first-stage classification results are output directly. The platform's online recognition interface and the deployment of the field monitoring equipment are shown in Fig. 11.

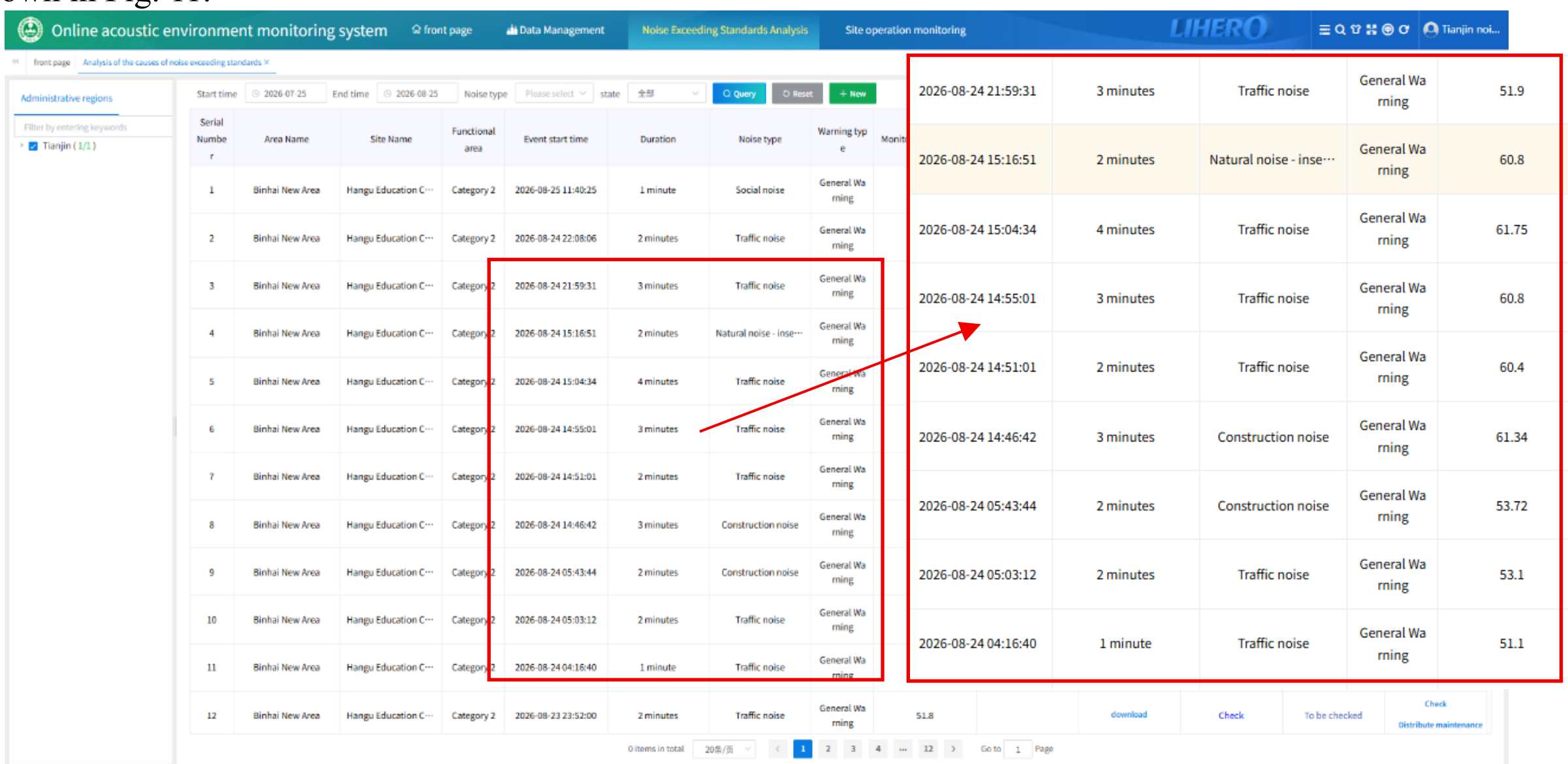

**(a)** Online classification platform

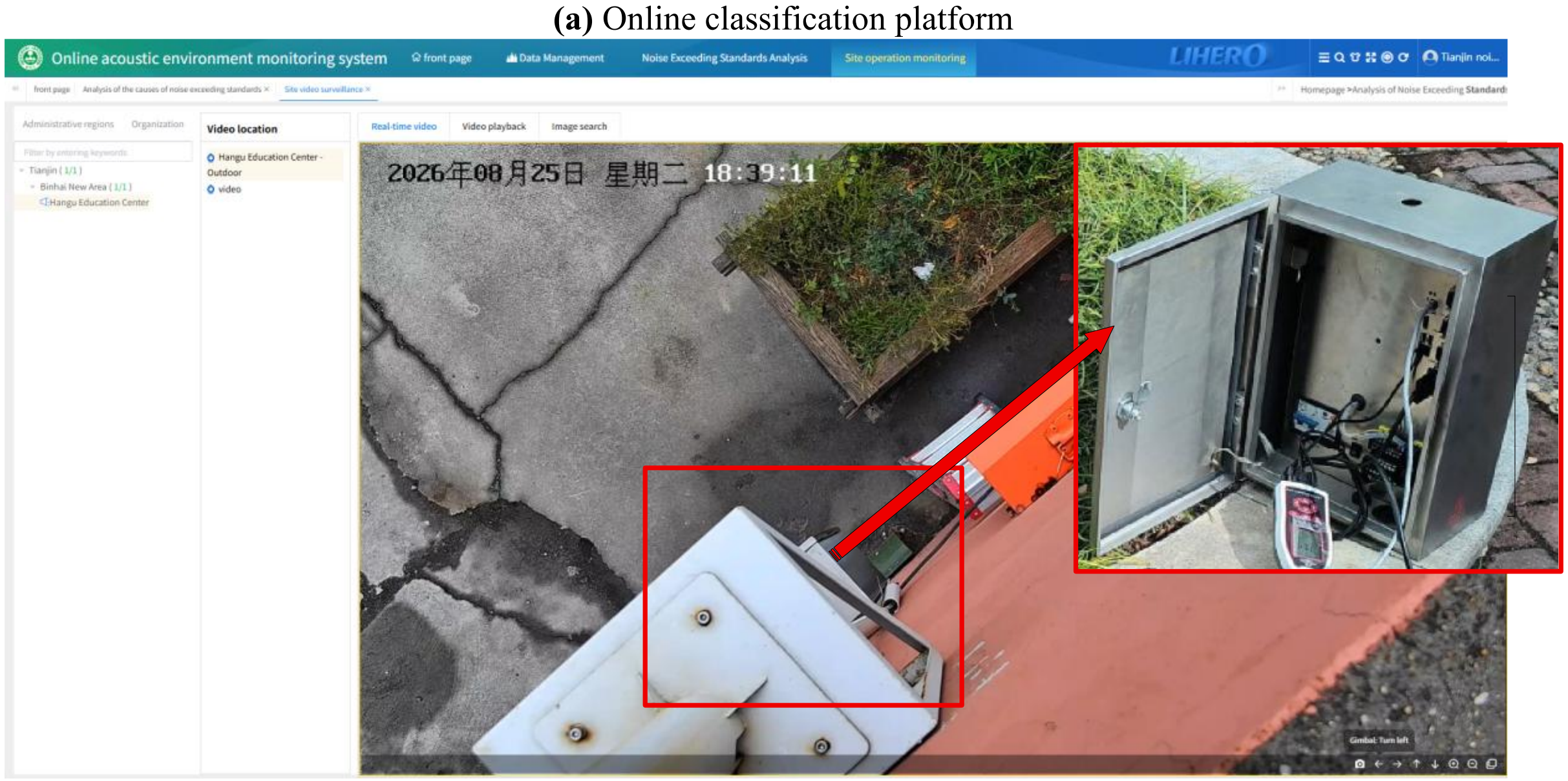

**(b)** Field scene and acoustic acquisition equipment

**Fig. 11.** Deployment of the online platform and field monitoring equipment.

In Fig.11(a), the platform's online recognition interface is shown and the system can record and display recognition results such as the sound category, duration, and early-warning information. In Fig.11(b), the actual monitoring environment and the deployment of the front-end acoustic acquisition equipment are shown.

To further evaluate the practical performance of the system under complex acoustic-scene conditions, the 1000 field-monitoring acoustic samples are randomly selected from data collected nearly five months at the Hangu Education

Center in Binhai New Area, Tianjin, China. After removing the 212 invalid recordings that containing unclassifiable noises, the 788 valid samples are retained for quantitative evaluation.

Unlike acoustic samples in a manually processing dataset, the recording sounds in field may simultaneously contain multiple acoustic sources and therefore may be associated with more than one valid label. Accordingly, a prediction is considered correct if the predicted label matches any one of the assigned truth labels. Because the class distribution in the field-monitoring dataset is highly imbalanced, weighted-average accuracy and weighted-average F1-score are used to evaluate the overall classification performance. The weight that assigned to each class is determined by its proportion of valid samples, i.e.,

$$w_i = \frac{n_i}{\sum_{k=1}^{K} n_k} \tag{20}$$

where $n_i$ is the number of valid samples in the *i-th* class and $K$ is the number of classes that containing valid samples. The classes that marked as N/A are excluded from the calculation. The weighted-average accuracy and the F1-score are calculated as

$$Accuracy_{weighted} = \sum_{i=1}^{K} w_i \times Accurcy_i \tag{21}$$

$$F1-score_{weighted} = \sum_{i=1}^{K} w_i \times F1-score_i \tag{22}$$

The highly imbalanced sample distribution in Table 6 reflects the actual acoustic characteristics of the deployment site. The Hangu Education Center is not located in an industrial zone; therefore, no valid C1 samples are recorded. The construction activities are also limited in the surrounding area, which results in only 18 C4 samples. During the monitoring period, no strong thunderstorm events occurred, and the site was not characterized by concentrated frog or insect activities. Wind sounds were also generally too weak to be captured as valid target events by the monitoring system. Consequently, no valid samples were obtained for S06, S07, S10, or S11. The uneven distribution therefore represents the actual composition and occurrence frequency of sound sources at the deployment site rather than an artificial imbalance introduced during dataset construction.

**Table 6.** Classification results on the field-monitoring dataset.

| Coarse Class | Fine class | Samples | Correct Samples | Incorrect Samples | Coarse | | Fine | | Overall Weighted | |
|---|---|---|---|---|---|---|---|---|---|---|
| | | | | | Accuracy | F1-score | Accuracy | F1-score | Accuracy | F1-score |
| C1 | -- | N/A | N/A | N/A | N/A | N/A | N/A | N/A | | |
| C2 | S06 | N/A | N/A | N/A | | | N/A | N/A | | |
| | S07 | N/A | N/A | N/A | | | N/A | N/A | | |
| | S08 | 263 | 209 | 54 | | | 0.9713 | 0.9807 | | |
| | S09 | 2 | 1 | 1 | 0.7963 | 0.8830 | 1.0000 | 1.0000 | | |
| | S10 | N/A | N/A | N/A | | | N/A | N/A | **0.8718** | **0.8740** |
| | S11 | N/A | N/A | N/A | | | N/A | N/A | | |
| | S12 | 5 | 5 | 0 | | | 1.0000 | 1.0000 | | |
| C3 | -- | 127 | 109 | 18 | 0.8583 | 0.8289 | N/A | N/A | | |
| C4 | -- | 18 | 12 | 6 | 0.6667 | 0.7742 | N/A | N/A | | |
| C5 | -- | 373 | 357 | 16 | 0.9571 | 0.8981 | N/A | N/A | | |
| Weighted Average | **--** | **788** | **693** | **95** | **0.8794** | **0.8789** | 0.9721 | 0.9812 | -- | |
| C2 end-to-end fine-grained classification | | | | | | | **0.7741** | **0.8685** | -- | |

At the coarse-grained classification stage, 693 of the 788 valid samples are classified correctly under the one-match criterion, resulting in an overall accuracy of 87.94% and a weighted-average F1-score of 87.89%. The classification accuracies for C2, C3, C4, and C5 are 79.63%, 85.83%, 66.67%, and 95.71%, respectively. These results demonstrate that this proposed method maintains good overall classification performance under the highly imbalanced class distribution encountered in the real-world monitoring environment.

Within C2, the 215 samples are correctly identified at the coarse-grained stage, and subsequently passed to the natural-sound fine-grained classifier. The fine-grained classifier achieved a weighted-average accuracy of 97.21% and a weighted-average F1-score of 98.12%. For S08, the accuracy and the F1-score are 97.13% and 98.07%, respectively. All routed samples belonging to S09 and S12 are correctly classified, with both classes achieving the accuracy and the F1-score of 100.00%. When both coarse-grained routing and subsequent fine-grained recognition are considered, the end-to-end accuracy and the F1-score for C2 are 77.41% and 86.85%, respectively. Across all 788 field-monitoring samples, the complete deployed classification architecture achieved an overall weighted accuracy of 87.18% and a weighted-average F1-score of 87.40%. These results verify the effectiveness of the proposed hierarchical procedure in performing both coarse-grained recognition and within-category fine-grained recognition on real-world field-monitoring data.

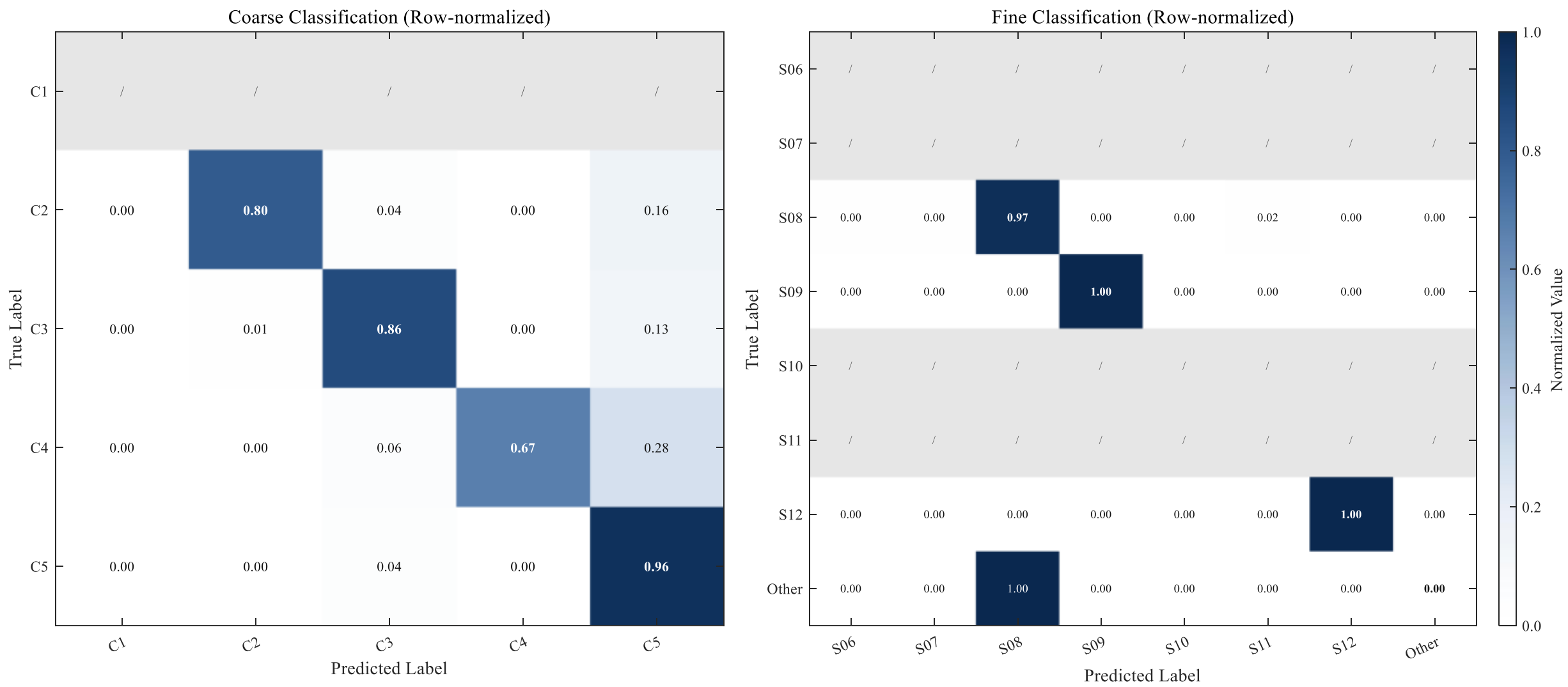

**Fig. 12.** Row-normalized confusion matrices for field-test classification.

As shown in Fig. 12, the predictions of both classification stages are generally concentrated along the main diagonal, which indicates that most field samples are correctly recognized. The remaining errors are confined to a small number of coarse-grained categories and natural-sound subclasses, which demonstrates the effectiveness and practical feasibility of the proposed method for urban acoustic environment monitoring.

## 5. Conclusions

This paper focuses on the problems of time-frequency feature modeling and imbalanced multiclass classification of environmental sounds in complex acoustic scenes, and proposes an environmental noise recognition method based on the FTANet architecture and a hierarchical strategy. A lightweight Frequency-Time Adaptive Network (FTANet) was designed to model time-domain dynamics and frequency-band distributions by separately using asymmetric one-dimensional convolutions. Adaptive fusion attention was further employed to perform channel-wise recalibration of the fused features, thereby enhancing the representation of discriminative acoustic information.

This two-stage hierarchical classification method is also constructed, which consists of coarse-grained category discrimination and within-category fine-grained recognition. On the self-constructed environmental sound dataset with five coarse-grained categories, the FTANet achieves an accuracy of 95.45% and a Macro-F1 score of 95.23%, with only 0.2004 M parameters and 0.3011 G FLOPs. The FTANet achieves the highest performance among the convolution-based models and ranks second only to AST. Compared with the AST, the FTANet reduces the parameter count and FLOPs by approximately 99.76% and 99.41%, respectively, which demonstrates a favorable balance between classification performance and model complexity. On UrbanSound8K, the FTANet achieves a mean accuracy of 79.24% and a mean F1-score of 80.37% across the ten folds, which outperforms all the convolution-based comparison models and validates its generalization performance across different datasets. For the 25-class environmental sound classification task, the two-stage hierarchical method improves the overall accuracy from 71.32% to 81.49% and the Macro-F1 score from 68.51% to 79.58%, with improvements of 10.17% and 11.07%, respectively. These results demonstrate that it can effectively reduce intercategory feature interferences and improve fine-grained recognition performances by decomposing the global classification task into coarse-grained category discrimination and within-category fine-grained recognition.

The proposed method is further integrated into the Tianjin Online Acoustic Environment Monitoring Platform, where the hierarchical classification procedure and the second-stage sub-classifier for natural sounds are deployed. Field testing is conducted on 788 valid acoustic samples that obtained from nearly five months of monitoring data. Despite the highly imbalanced samples, the site-specific class distribution and the multiple sound sources, the overall weighted accuracy of this method is 87.18% and its weighted-average F1-score is 87.40%. These results verify the effectiveness and practical feasibility of the proposed hierarchical environmental noise recognition method for the urban acoustic environment monitoring.

## References

[1] CARRATÙ M, LIGUORI C, PACIELLO V, et al. A Sound Level Meter featured with automatic estimation of the measurement uncertainty[J]. Measurement, 2022, 188: 110409.

[2] ZAGUBIEŃ A, WOLNIEWICZ K. Development and validation of a portable test stand for sound measurement near the building façade[J]. Measurement, 2023, 214: 112856.

[3] BANSAL A, GARG N K. Environmental sound classification: A descriptive review of the literature[J]. Intelligent Systems with Applications, 2022, 16: 200115.

[4] TAN J K A, HASEGAWA Y, LAU S K. A comprehensive environmental sound categorization scheme of an urban city[J]. Applied Acoustics, 2022, 199: 109018.

[5] AUMOND P, LAVANDIER C, RIBEIRO C, et al. A study of the accuracy of mobile technology for measuring urban noise pollution in large scale participatory sensing campaigns[J]. Applied Acoustics, 2017, 117: 219-226.

[6] CAO J, CAO M, WANG J, et al. Urban noise recognition with convolutional neural network[J]. Multimedia Tools and Applications, 2019, 78(20): 29021-29041.

[7] İŞLER B. Urban sound recognition in smart cities using an IoT-fog computing framework and deep learning models: A performance comparison[J]. Applied Sciences, 2025, 15(3): 1201.

[8] HUANG Z, LIU C, FEI H, et al. Urban sound classification based on 2-order dense convolutional network using dual features[J]. Applied Acoustics, 2020, 164: 107243.

[9] XIE J, ZHONG Y, ZHANG J, et al. A review of automatic recognition technology for bird vocalizations in the deep learning era[J]. Ecological Informatics, 2023, 73: 101927.

[10] RADHAKRISHNAN R, DIVAKARAN A, SMARAGDIS P. Audio analysis for surveillance applications[C]//Proceedings of the 2005 IEEE Workshop on Applications of Signal Processing to Audio and Acoustics. Piscataway: IEEE, 2005: 158-161.

[11] CROCCO M, CRISTANI M, TRUCCO A, et al. Audio surveillance: A systematic review[J]. ACM Computing Surveys, 2016, 48(4): 52:1-52:46.

[12] LAFFITTE P, WANG Y, SODOYER D, et al. Assessing the performances of different neural network architectures for the detection of screams and shouts in public transportation[J]. Expert Systems with Applications, 2019, 117: 29-41.

[13] MILIVOJČEVIĆ M, ĆIRIĆ D, PREZELJ J, et al. Analysis of unsupervised learning approach for classification of vehicle fuel type using psychoacoustic features[J]. Measurement, 2024, 227: 114318.

[14] LV M, SUN Z, ZHANG M, et al. Sound recognition method for white feather broilers based on spectrogram features and the fusion classification model[J]. Measurement, 2023, 222: 113696.

[15] WANG W, GAO Y. Pipeline leak detection method based on acoustic-pressure information fusion[J]. Measurement, 2023, 212: 112691.

[16] DONG X, YIN B, CONG Y, et al. Environment sound event classification with a two-stream convolutional neural network[J]. IEEE Access, 2020, 8: 125714-125721.

[17] HERSHEY S, CHAUDHURI S, ELLIS D P W, et al. CNN architectures for large-scale audio classification[C]//Proceedings of the 2017 IEEE International Conference on Acoustics, Speech and Signal Processing. Piscataway: IEEE, 2017: 131-135.

[18] İNİK Ö. CNN hyper-parameter optimization for environmental sound classification[J]. Applied Acoustics, 2023, 202: 109168.

[19] CINAR I. DeepSoundVisionNet: A new approach to urban sound classification using visual representations of audio signals[J]. Engineering Applications of Artificial Intelligence, 2026, 170: 114193.

[20] LI J, HAN L, LI X, et al. An evaluation of deep neural network models for music classification using spectrograms[J]. Multimedia Tools and Applications, 2022, 81(4): 4621-4647.

[21] WANG H, ZOU Y, CHONG D, et al. Environmental sound classification with parallel temporal-spectral attention[C]//Proceedings of Interspeech 2020. [S.l.]: ISCA, 2020: 821-825.

[22] LI D, LI C, YANG J, et al. Bayesian optimization-attention-feedforward neural network based train traction motor-gearbox coupled noise prediction[J]. Measurement, 2024, 238: 115323.

[23] LI Z, OUYANG B, XU X, et al. Non-stationary mechanical sound source separation: An all-neural beamforming network driven by time–frequency convolution and self-attention[J]. Measurement, 2025, 242: 115933.

[24] LIU Y, XU Z, HE Y, et al. Acoustic fault diagnosis method for rotating machinery based on improved spectral subtraction and CNN-TCN model[J]. Measurement, 2025, 256: 118482.

[25] DAI W, DAI C, QU S, et al. Very deep convolutional neural networks for raw waveforms[C]//Proceedings of the 2017 IEEE International Conference on Acoustics, Speech and Signal Processing. Piscataway: IEEE, 2017: 421-425.

[26] VALERO X, ALÍAS F. Hierarchical classification of environmental noise sources considering the acoustic signature of vehicle pass-bys[J]. Archives of Acoustics, 2012, 37(4): 423-434.

[27] GUZHOV A, RAUE F, HEES J, et al. ESResNet: Environmental sound classification based on visual domain models[C]//Proceedings of the 2020 25th International Conference on Pattern Recognition. Piscataway: IEEE, 2021: 4933-4940.

[28] GONG Y, CHUNG Y A, GLASS J. AST: Audio spectrogram transformer[C]//Proceedings of Interspeech 2021. [S.l.]: ISCA, 2021: 571-575.
[29] RIAZATI SERESHT H, MOHAMMADI K. Environmental sound classification with low-complexity convolutional neural network empowered by sparse salient region pooling[J]. IEEE Access, 2023, 11: 849-862.
[30] RANMAL D, RANASINGHE P, PARANAYAPA T, et al. ESC-NAS: Environment sound classification using hardware-aware neural architecture search for the edge[J]. Sensors, 2024, 24(12): 3749.
[31] SALAMON J, JACOBY C, BELLO J P. A dataset and taxonomy for urban sound research[C]//Proceedings of the 22nd ACM International Conference on Multimedia. New York: ACM, 2014: 1041-1044.